%% file: top.tex
\documentclass[runningheads]{llncs}
\usepackage{graphicx}
\usepackage{comment}
\usepackage{amsmath,amssymb} 
\usepackage{color}
\usepackage{overpic}
\usepackage{xcolor}
\usepackage{bm}
\usepackage{caption}

\usepackage[width=122mm,left=12mm,paperwidth=146mm,height=193mm,top=12mm,paperheight=217mm]{geometry}

\newcommand{\etal}[0]{\textit{et al.}}
\newcommand{\jmb}[1]{{\bm{#1}}}

\begin{document}
\pagestyle{headings}
\mainmatter
\def\ECCVSubNumber{2805}  


\title{Conditional Entropy Coding for Efficient Video Compression} 

%
\author{Jerry Liu \and Shenlong Wang \and Wei-Chiu Ma \and Meet Shah \and Rui Hu \and Pranaab Dhawan \and Raquel Urtasun}
%
%
\institute{Uber ATG \\
\email{\{jerryl, slwang, weichiu, meet.shah, rui.hu, pdhawan, urtasun\}@uber.com}}

\maketitle

\begin{abstract}

We propose a very simple and efficient video compression framework that only focuses on modeling the conditional entropy between frames. Unlike prior learning-based approaches, we reduce complexity by not performing any form of explicit transformations between frames and assume each frame is encoded with an independent state-of-the-art deep image compressor. 
We first show that a simple architecture modeling the entropy between the image latent codes is as competitive as other neural video compression works and video codecs while being much faster and easier to implement.
We then propose a novel internal learning extension on top of this architecture that brings an additional $\sim 10\%$ bitrate savings without trading off decoding speed.
Importantly, we show that our approach 
outperforms H.265 and other deep learning baselines in MS-SSIM on higher bitrate UVG video, and against all video codecs on lower framerates, while being thousands of times faster in decoding than deep models utilizing an autoregressive entropy model. 
\end{abstract}

\input{introduction.tex}
\input{related_work.tex}

\input{method_1.tex}

\input{experiments.tex}

\input{conclusion.tex}



\clearpage
%
%
\bibliographystyle{splncs04}
\bibliography{egbib}

\clearpage
\input{supp.tex}

\end{document}

%% file: introduction.tex
\section{Introduction}

The efficient storage of video data is vitally important to an enormous number of settings, from online websites such as Youtube and Facebook to robotics settings such as drones and self-driving cars. This necessitates the use of good video compression algorithms. Both image and video compression are fields that have been extensively researched in the past few decades. Traditional image codecs such as JPEG2000, BPG, and WebP, and traditional video codecs such as HEVC.H.265, AVC/H.264 \cite{wiegand_h264,hevc} are well-known and have been widely used. They are hand-engineered to work well in a variety of settings, but the lack of learning involved in the algorithm leaves room open for more end-to-end optimized solutions.


Recently, there has been an explosion of deep-learning based image compressors that have been demonstrated to outperform BPG on a variety of evaluation datasets across both MS-SSIM and PSNR as evaluation metrics \cite{liu_neuralvid,minnen_jointpriors,balle_varhyperprior,liu_dsic}.
This explosion has also recently happened in video compression on a somewhat smaller scale, with the latest advances being able to outperform H.265 on MS-SSIM and PSNR in certain cases \cite{rippel_learnedvidcomp,habibian_rdauto,djelouah_interframe,lu_dvc}. 
Many of these approaches \cite{wu_vidinterpolation,rippel_learnedvidcomp,djelouah_interframe} involve learning-based generalizations of the traditional video compression techniques of motion-compensation, 
frame interpolation and residual coding. 

\begin{figure}
	\centering
	\includegraphics[width=0.47 \linewidth]{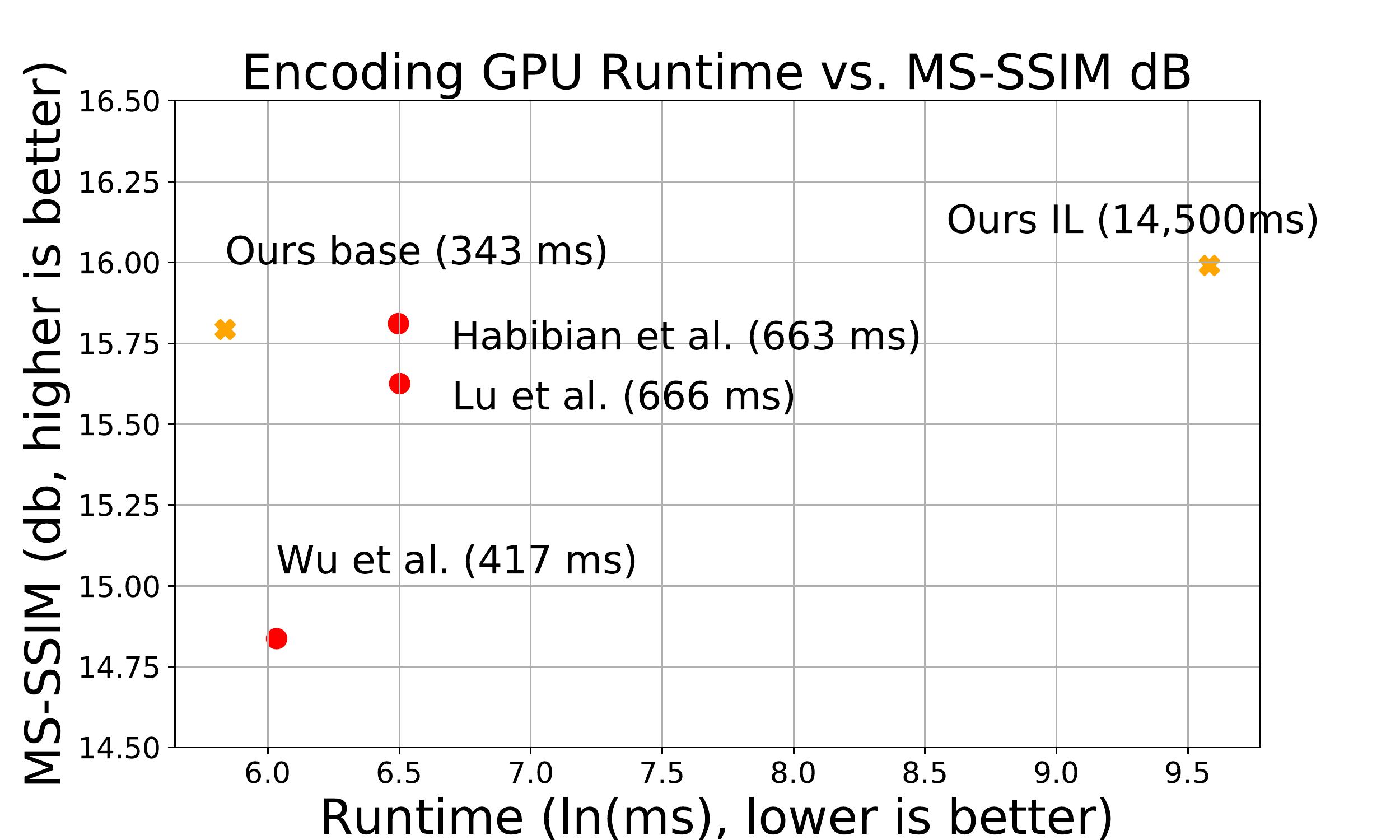}
	\includegraphics[width=0.47 \linewidth]{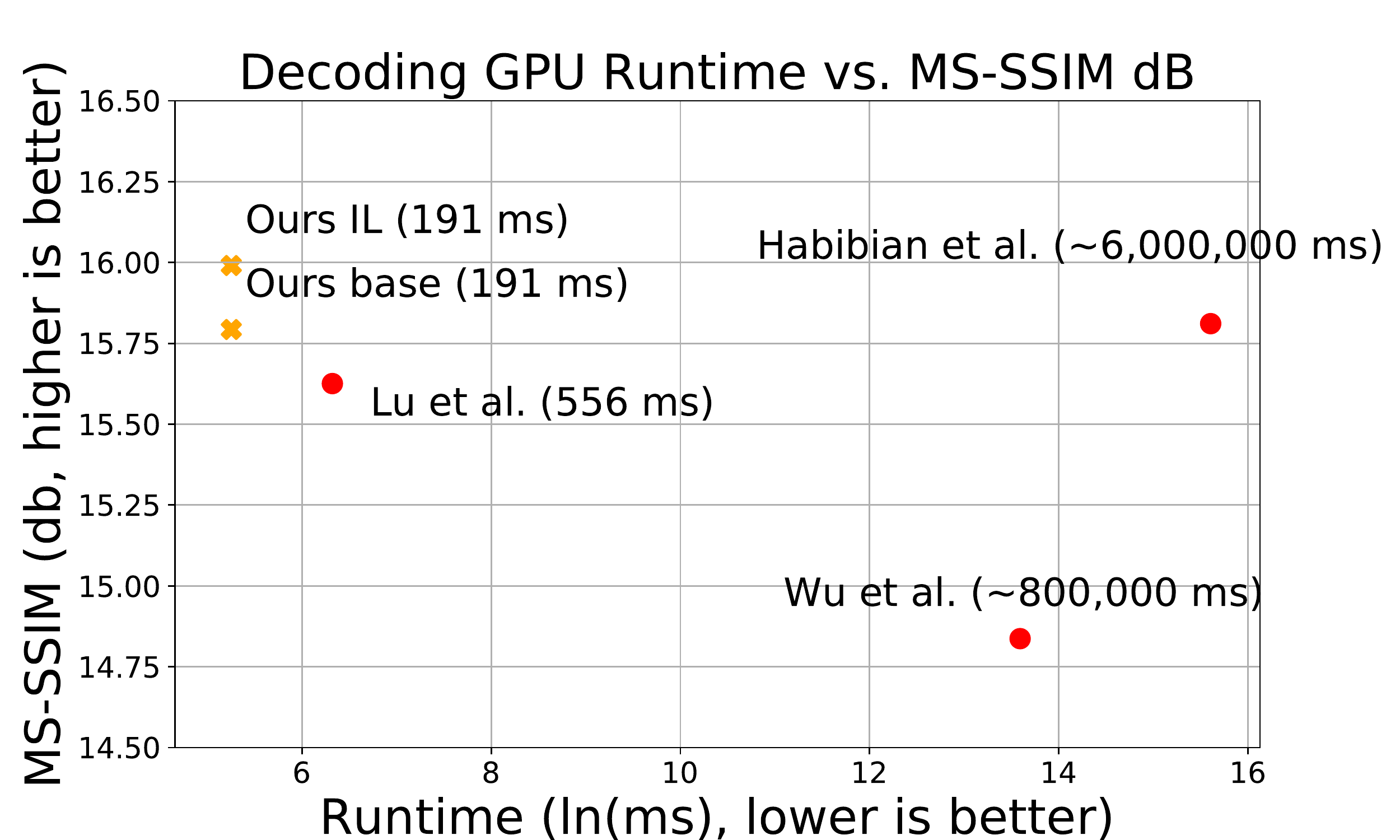}
	\caption{Plots indicating the GPU runtime vs. MS-SSIM of our model vs. other deep approaches at bitrate 0.2, averaged over a 1920 $\times$ 1080 UVG frame. Runtimes are shown independent of the entropy coding implementation. We interpolate to obtain the MS-SSIM estimate at the exact bitrate. 
	}
	\label{fig:comp_runtime_figure}
	\vspace{-4mm}
\end{figure}

While achieving impressive distortion-rate curves, there are several major facts blocking the wide adoption of these approaches for real-world, generic video compression tasks. First, most aforementioned approaches are still slower than standard video codecs at both encoding and decoding stage; moreover, due to the  the fact that they explicitly perform interpolation and residual coding between frames, a majority of the computations cannot be parallelized to accelerate coding speed; finally, the domain bias of the training dataset makes it difficult to generalize well to a wide range of different type of videos.

\vspace{-3mm}
\begin{figure}
	\centering
	\includegraphics[width=\linewidth]{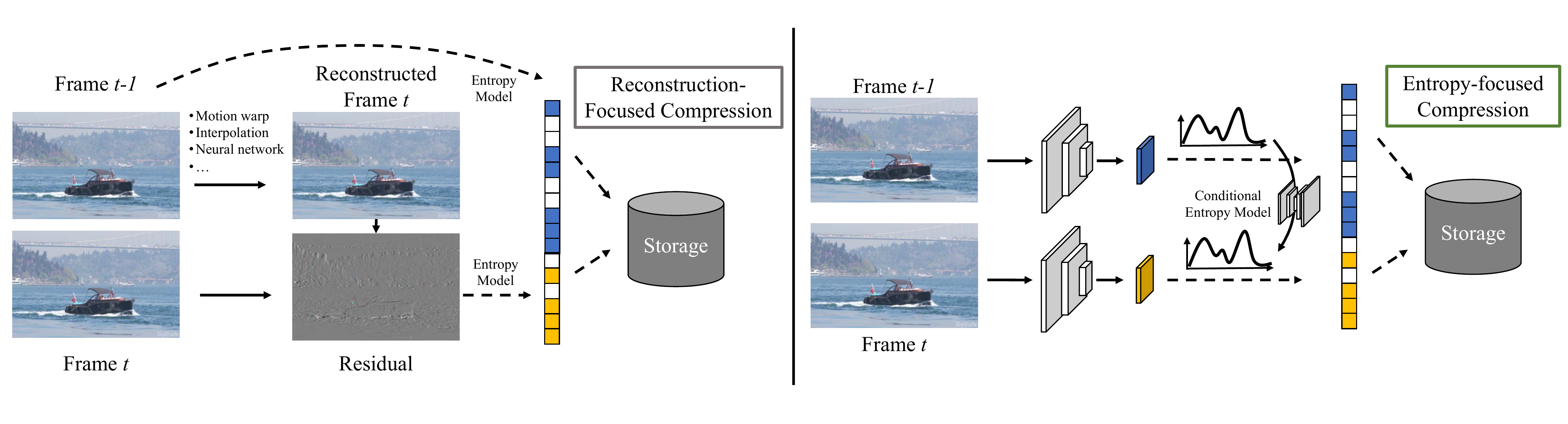}
	\caption{An illustration of the explicit transformations used in removing redundant information in subsequent frames vs. probabilistic modeling used in entropy coding. A typical lossy compression algorithm will contain elements of both approaches.
	}
	\label{fig:explicit_implicit}
\end{figure}

\begin{figure*}
	\centering
	\includegraphics[width=\linewidth]{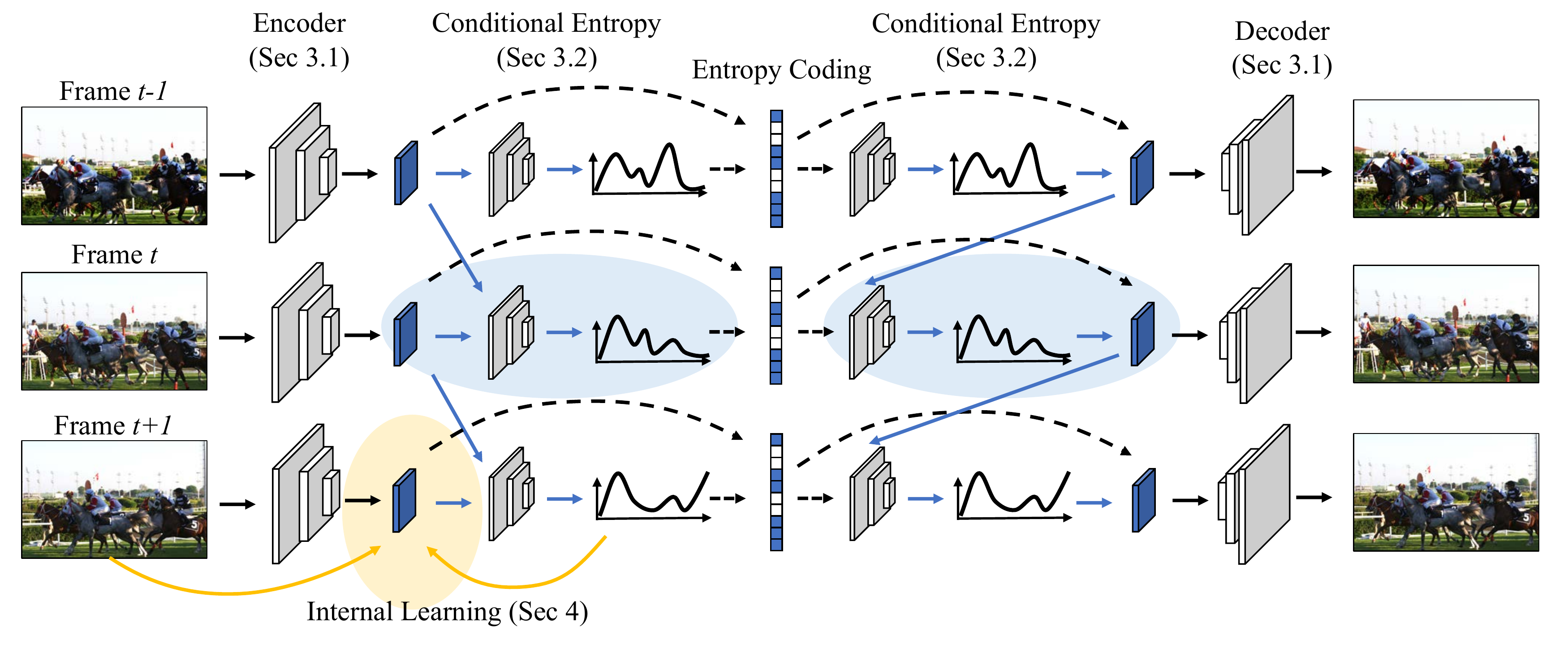}
	\caption{Overview of the architecture of our approach. We highlight our key contribution, namely the conditional entropy model and internal learning in blue and yellow, respectively. }
	\label{fig:architecture}
	\vspace{-2mm}
\end{figure*}

In this paper, we address these issues by creating a remarkably simple \textit{entropy-focused} video compression approach that is not only competitive with prior state-of-the-art learned compression, but also \textit{significantly faster} (see Fig. \ref{fig:comp_runtime_figure}), rendering it a practical alternative to existing video codecs. Such an entropy-focused approach focuses on better capturing the correlations between frames during entropy coding rather than performing explicit transformations (e.g. motion compensation). 
Our contributions are two-fold (illustrated in Fig. \ref{fig:architecture}). First, we propose a base model consisting only of a \textbf{conditional entropy model} fitted on top of the latent codes produced by a deep single-image compressor. The intuition for why we don't need explicit transformations can be visualized in Fig. \ref{fig:explicit_implicit}: given two video frames, prior works would code the first frame to store the full frame information while coding the second frame to store explicit motion information from frame 1 as well as residual bits. On the other hand, our approach encodes both frames as independent image codes, and reduces the joint bitrate  by fitting probability model (an \textit{entropy} model) to maximize the probability of the second image code given the first. We can thus extend this to a full video sequence by still encoding every frame independently, and simply considering every adjacent pair of frames for the probability model. 
While entropy modeling has been a subcomponent of prior works \cite{rippel_learnedvidcomp,habibian_rdauto,lu_dvc,wu_vidinterpolation,lee_contextadapt}, they have tended to be very simple \cite{rippel_learnedvidcomp}, only dependent on the image itself \cite{lu_dvc,djelouah_interframe}, or use costly autoregressive models that are intractably expensive during decoding \cite{habibian_rdauto,wu_vidinterpolation}; here our conditional entropy model provides a viable means for video compression purely within itself. 

Our second contribution is to propose \textbf{internal learning} of the latent code during inference. Prior works in video compression operate by using a fixed encoder during the inference/encoding stage. As a result, the latent codes of the video is not optimized towards reconstruction/entropy estimation for the specific test video. We observe as long as the decoder is fixed, we can trade off \textit{encoding runtime} to further optimize the latent codes along the rate-distortion curve, while not affecting \textit{decoding runtime} (Fig. \ref{fig:comp_runtime_figure}, right).

We validate the performance of the proposed approach over several datasets across various framerates. We show that at standard framerates, our base model is much faster and easier to implement than most state-of-the-art deep video benchmarks, while matching or outperforming these benchmarks as well as H.265 on MS-SSIM. Adding internal learning provides additional $\sim 10\%$ bitrate gains with the same decoding time. Additionally, on lower framerates, our models outperform H.265 by a wide margin at higher bitrates. 
The simplicity of our method indicates that it is a powerful approach that is widely applicable across videos spanning a broad range of content, framerates, and motion. 

%% file: related_work.tex
\section{Background and Related Work}



\subsection{Deep Image Compression} 
There is an abundance of work on learned, lossy image-compression \cite{toderici_varimgcomp,toderici_fullimgcomp,balle_imgcomp_end2end,balle_varhyperprior,theis_imgcomp_ae,mentzer_condprobimg,rippel_rtimgcomp,minnen_jointpriors}. In general, these works follow a general autoencoder architecture minimizing the rate-distortion tradeoff. Typically, an encoder transforms the image into a latent space, quantizes the symbols, and applies entropy coding (typically arithmetic/range coding) on the symbols to output a compressed bitstream. During decoding, the recovered symbols are then fed through a decoder for  image reconstruction. 

Recent works approximate the rate-distortion tradeoff $\ell(\jmb{x}, \hat{\jmb{x}})  + \beta R(\hat{\jmb{y}})$ in a differentiable manner by replacing the bitrate term $R$ with the cross-entropy between the code distribution and a learned ``prior" probability model: $R \approx \mathbb{E}_{x \sim p_{data}}[\log p(E(\jmb{x}); \jmb{\theta})]$. Shannon's source coding theorem \cite{shannon_lowerbound} indicates that the bitrate can asymptotically approach, but can never be lower than, the cross-entropy. One way to achieve this optimal bitrate during entropy coding is to use the learned ``'prior" model as the probability map during arithmetic coding or range coding to code the symbols. Hence, the smaller the cross-entropy term, the more the bitrate can be reduced. This then implies that the more expressive the prior model in modeling the true distribution of latent codes, the smaller the overall bitrate. 

Sophisticated prior models have been designed for the quantized representation in order to minimize the cross-entropy with the code distribution. 
Autoregressive models \cite{minnen_jointpriors,mentzer_condprobimg,toderici_fullimgcomp}, hyperprior models \cite{balle_varhyperprior,minnen_jointpriors}, and factorized models \cite{theis_imgcomp_ae,balle_imgcomp_end2end,balle_varhyperprior} have been used to model this prior. \cite{minnen_jointpriors} and \cite{mentzer_lossless} suggest that using an autoregressive model is intractably slow in practice, as it requires a pass through the model for every single pixel during decoding. \cite{mentzer_lossless} suggests that the hyperprior approach presents a good tradeoff between speed and performance.


A recent model by Liu \etal \cite{liu_nonlocalcomp} presents an extension of \cite{minnen_jointpriors} using residual blocks in the encoder/decoder, outperforming BPG and other deep models on both PSNR and MS-SSIM on Kodak. 

\subsection{Video Compression} \label{sec:old_multiview}
Conceptually, traditional video codecs such as H.264 / H.265 exploit temporal correlations between frames by categorizing the frames as follows \cite{hevc,wiegand_h264}:
\begin{itemize}
	\item I-frames: compressed as an independent image
	\item P-frames: predicted from past frames using block-based flow estimate, then encode residual.
	\item B-frames: similar to P-frames but predicted from both past and future frames.
\end{itemize}

In order to predict P/B-frames, the motion between frames is predicted via block matching (and the flow is uniformly applied within blocks), and then the resulting difference is separately encoded as the "residual." Generally, if neighboring frames are temporally correlated, encoding the motion and residual vectors requires fewer bits than recording the subsequent frame independently. 
 
Recently, several deep-learning based video compression frameworks  \cite{wu_vidinterpolation,rippel_learnedvidcomp,liu_neuralvid,lu_dvc,djelouah_interframe,han_deepgenvc} have been developed.  Both Wu \etal \cite{wu_vidinterpolation} and Lu \etal\cite{lu_dvc} attempt to generalize various parts of the motion-compensation and residual learning framework with neural networks, and get close to H.265 performance (on \textit{verfast} setting).
Rippel~\etal \cite{rippel_learnedvidcomp} were able to achieve state-of-the-art results in both MS-SSIM compared to H.265 by generalizing flow/residual coding with a global state, spatial multiplexing, and adaptive codelength regularization. Djelouah \etal \cite{djelouah_interframe} jointly decode motion and blending coefficients from references frames, and represent residuals in latent space. Habibian \etal \cite{habibian_rdauto} utilizes a 3D convolutional architecture to avoid motion compensation, as well as an autoregressive entropy model to outperform \cite{wu_vidinterpolation,lu_dvc}. 

These prior works generally require specialized modules and explicit transformations, with the entropy model being an oftentimes intractable autoregressive subcomponent \cite{wu_vidinterpolation,djelouah_interframe,habibian_rdauto}. A more closely related work is that of Han et al. \cite{han_deepgenvc}, who propose to model the entropy dependence between codes with an LSTM: $p(\jmb{y}_{i} | \jmb{y}_{< i})$. In contrast to these prior works, we focus on a \textit{entropy-only} approach, with no explicit transformations across time. More importantly, our base approach carefully exploits the parallel nature of frame encoding/decoding, rendering it orders of magnitude faster than other state-of-the-art while being just as competitive.



\subsection{Internal Learning} 
The concept of internal learning is not new. It is similar to the sample-specific nature of transductive learning \cite{vapnik_naturestat,sun_testtime}. 
``Internal Learning" is a term proposed in \cite{shocher_zeroshotinternal,glasner_superres}, which exploits the internal recurrence of information within a single-image to train an unsupervised super-resolution algorithm. Many related works have also trained deep networks on a single example, from DIP \cite{ulyanov_dip} to GANs \cite{shocher_ingan,shaham_singan,zhou_nonstattex}. Also related is Sun et al. \cite{sun_testtime} who propose ``test-time training" on an auxiliary function for each test instance on supervised classification tasks. Concurrently and independently from our work, Campos et al. propose context adaptive optimization in image compression \cite{campos_cao}, which has demonstrated promising results on finetuning each latent code towards its test image. 

In our setting, we leverage the fact that in video compression the ground-truth is simply the video itself, and we apply internal learning in a way that obeys codebook consistency while decreasing the conditional entropy between video frames during decoding. There are unique advantages to using internal learning in our entropy-only video compression setting: it can optimize for conditional entropy between codes in a way that an independent frame encoder cannot (see Section \ref{sec:test_time}).

%% file: method_1.tex
\section{Entropy-focused Video Compression}



Our base model consists of two components: we first encode each frame $\jmb{x}_i$ of a video $\jmb{x}$ with a straightforward, off-the-shelf image compressor consisting of a deep image encoder/decoder (Section \ref{sec:image_comp}) to obtain discrete image codes $\jmb{y}_i$. Then, we capture the temporal relationships between our $\jmb{y}_i$'s with a  conditional entropy model that approximates the joint entropy of the video sequence 
(Section \ref{sec:cond_entropy}). 
The model is trained end-to-end with respect to the rate-distortion loss function (Section \ref{sec:rate_distortion_loss}).


\subsection{Single-image Encoder/Decoder} \label{sec:image_comp}
We encode every video frame $\jmb{x}_i$ separately with a deep image compressor into a quantized latent code $\jmb{y}_i$; note that each $\jmb{y}_i$ contains \textit{full} information to reconstruct each frame $i$ and does not depend on previous frames. Our choice of architecture for single-image compression borrows heavily from the state-of-the-art model presented by Liu et al. \cite{liu_nonlocalcomp}, which has shown to outperform BPG on both MS-SSIM and PSNR. The architecture consists of the image encoder, quantizer, and image decoder.
We simplify the model in two ways compared to the original paper: we remove all non-local layers for efficiency/memory reasons, and we remove the autoregressive context estimation due to its decoding intractability (\cite{minnen_jointpriors,mentzer_condprobimg}, also see Fig. \ref{fig:comp_runtime_figure}).


More details about the image encoder/decoder architecture are found in supplementary material. In our video compression model, we use the image encoder/quantizer to produce the quantized code $\jmb{y}_i$, and the image decoder to produce the reconstruction $\jmb{\hat{x}}_i$. We do not use the existing entropy model (inspired from \cite{balle_varhyperprior}, \cite{minnen_jointpriors}) which are only designed for modeling the intra-image entropy; instead we design our own conditional entropy model, as detailed next. 

\subsection{Conditional Entropy Model for Video Encoding} \label{sec:cond_entropy}



Our \textit{entropy model} models 
the joint entropy of the video frame codes 
with a deep network in order to reduce the overall bitrate of the video sequence; this is because the cross-entropy between our entropy model and the actual code distribution is a tight lower bound of the bitrate \cite{shannon_lowerbound}. 
Our goal is to design our entropy model to capture the temporal correlations as well as possible between the frames such that it can  minimize the cross-entropy with the code distribution. Put another way, the bitrate for the entire video sequence code $R(\jmb{y})$ is tightly approximated by the cross-entropy between the code distribution induced by the encoder $\jmb{y} = E(\jmb{x}), \jmb{x} \sim p_{data}$ 
and our probability model $p(\cdot | \jmb{\theta})$: $\mathbb{E}_{x \sim p_{data}}[\log p(\jmb{y}; \jmb{\theta})]$. 

\begin{figure}
	\centering
	\includegraphics[width=1.0\linewidth]{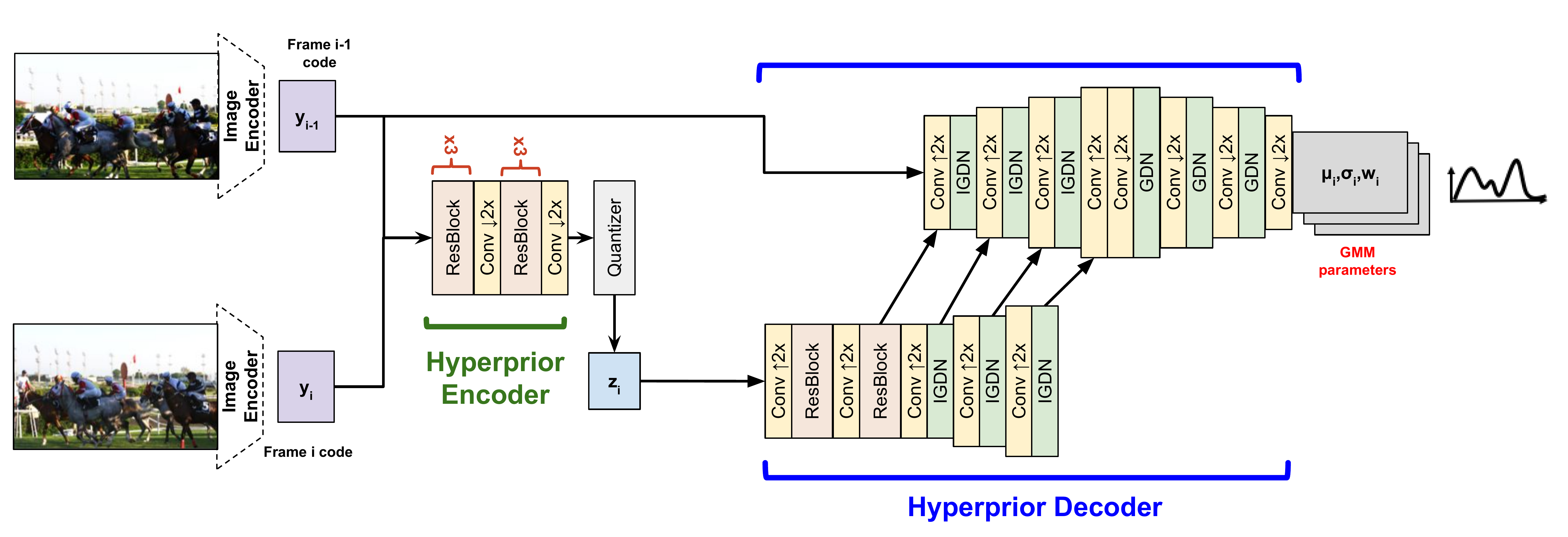}
	\caption{Diagram of our conditional entropy model, consisting of both a hyperprior encoder (top) and decoder (bottom).}
	\label{fig:cond_entropy_model}
	\vspace{-2mm}
\end{figure}

If $\jmb{y} = \{ \jmb{y}_1, \jmb{y}_2, ... \}$ represents the sequence of frame codes for the entire video sequence,
then a natural factorization of the joint probability $p(\jmb{y})$ would be to have every subsequent frame depend on the previous frames: 
\begin{equation}
\begin{split}
R(\jmb{y}) \geq \mathbb{E}_{x \sim p_{data}}[\sum^n_{i=0}{\log p(\jmb{y}; \jmb{\theta})}] =  \mathbb{E}_{x \sim p_{data}}[\sum^n_{i=0}{\log p(\jmb{y}_i | \jmb{y}_{< i}; \jmb{\theta})}] \\
\end{split}
\end{equation}
While other approaches (e.g. B-frames) model dependence in a hierarchical manner, our factorization makes sense in online and low-latency settings, where we want to decode frames sequentially. 
We further make a 1st-order Markov assumption such that
each frame $\jmb{y}_i$ only depends on the previous frame $\jmb{y}_{i-1}$ and a small hyperprior code $\jmb{z}_i$. Note that $\jmb{z}_i$ counts as side information, inspired from \cite{balle_varhyperprior}, and must also be counted in the bitstream. 
We encode it with a hyperprior encoder with $\jmb{y}_i$ and $\jmb{y}_{i-1}$ as input (see Fig. \ref{fig:cond_entropy_model}).
We thus have 
\[
 R(\jmb{y}) \geq \mathbb{E}_{x \sim p_{\jmb{x}}}[\sum^n_{i=0}{\log p(\jmb{y}_i | \jmb{y}_{i-1}, \jmb{z}_i; \jmb{\theta}) + \log p(\jmb{z}_i; \jmb{\theta})}] 
 \]

We assume that the hyperprior code distribution $p(\jmb{z}_i; \jmb{\theta})$ is modeled as a factorized distribution, $p(\jmb{z}_i; \jmb{\theta}) = \prod_{j}{p(z_{ij} | \jmb{\theta_{\jmb{z}}})}$, where $j$ represents each dimension of $\jmb{z}_i$. Since each $z_{ij}$ is a discrete value, we design each $p(z_{ij} | \jmb{\theta_{\jmb{z}}}) = c_j(z_{ij} + 0.5 ;\jmb{\theta_{\jmb{z}}}) - c_j(z_{ij} - 0.5;\jmb{\theta_{\jmb{z}}})$, where each $c_j( \cdot ;\jmb{\theta_{\jmb{z}}})$ is a cumulative density function (CDF) parametrized as a neural network similar to \cite{balle_varhyperprior}. 
In the meantime, we also model each $p(\jmb{y}_i | \jmb{y}_{i-1}, \jmb{z}_i; \jmb{\theta})$ as a conditional factorized distribution: $\prod_{j}{p(y_{ij} | \jmb{y}_{i-1}, \jmb{z}_i; \jmb{\theta})}$, with $p(y_{ij} | \jmb{y}_{i-1}, \jmb{z}_i; \jmb{\theta}) = g_j(y_{ij} + 0.5 | \jmb{y}_{i-1}, \jmb{z}_i; \jmb{\theta_y}) - g_j(y_{ij} - 0.5 | \jmb{y}_{i-1}, \jmb{z}_i; \jmb{\theta_y})$, where $g_j$ is modeled as the CDF of a Gaussian mixture model:
$\sum_{k}{w_{jk}\mathcal{N}(\mu_{jk}, \sigma^{2}_{jk})}$. 
$w_{jk}$, $\mu_{jk}, \sigma_{jk}$ are all learned parameters depending on $\jmb{y}_{i-1}, \jmb{z}_i; \jmb{\theta_y}$. Similar to \cite{balle_varhyperprior,minnen_jointpriors,liu_dsic}, the GMM parameters are outputs of a deep hyperprior decoder. 

Note that our entropy model is \textit{not} autoregressive either at the pixel level or the frame level - mixture parameters for each latent ``pixel" $y_{ij}$ are predicted independently given $\jmb{y}_{i-1}, \jmb{z}_i$, hence requiring only one GPU pass per frame during decoding. Also, all $\jmb{y}_i$'s are produced independently with our image encoder, removing the need to specify keyframes. All these aspects are advantageous in designing a fast, online video compressor. Yet we also aim to make our model expressive such that our prediction for each pixel $p(y_{ij} | \jmb{y}_{i-1}, \jmb{z}_i; \jmb{\theta})$ can incorporate both local and global structure information surrounding that pixel. 

We illustrate this architecture in Fig. \ref{fig:cond_entropy_model}. Our hyperprior encoder encodes our hyperprior code $\jmb{z}_i$ as side information given $\jmb{y}_i$ and $\jmb{y}_{i-1}$ as input. Then, our hyperprior decoder takes $\jmb{z}_i$ and $\jmb{y}_{i-1}$ as input to predict the Gaussian mixture parameters for $\jmb{y}_i$: $\jmb{\sigma}_i$, $\jmb{\mu}_i$, and $\jmb{w}_i$. We can effectively think of $\jmb{z}_i$ as providing supplemental information to $\jmb{y}_{i-1}$ to better predict $\jmb{y}_i$. The hyperprior decoder first upsamples $\jmb{z}_i$ to the spatial resolution of $\jmb{y}_{i-1}$ with residual blocks; then, it uses deconvolutions and IGDN nonlinearities \cite{balle_gdn} to progressively upsample both $\jmb{y}_{i-1}$ and $\jmb{z}_{i}$ to different resolution feature maps, and fuses the $\jmb{z}_{i}$ feature to the $\jmb{y}_{i-1}$ at each corresponding upsampled resolution. This helps to incorporate changes between $\jmb{y}_{i-1}$ to $\jmb{y}_{i}$, encapsulated by $\jmb{z}_i$, at multiple resolution levels from more global features at the lower resolution to finer features at higher resolutions. Then, downsampling convolutions and GDN nonlinearities are applied to match the original spatial resolution of the image code and produce the mixture parameters for each pixel of the code. 

\subsection{Rate-distortion Loss Function} \label{sec:rate_distortion_loss}

We train our base compression models end-to-end to minimize the rate-distortion tradeoff objective used for lossy compression:
\begin{equation} \label{eq:rate_distortion_loss}
L(\jmb{x}) = \underbrace{\mathbb{E}_{x \sim p_{data}}[\sum^n_{i=0}{||\jmb{x}_i - \jmb{\hat{x}}_i||^2}]}_{\text{Distortion}}+
\lambda  \underbrace{\mathbb{E}_{x \sim p_{data}}[\sum^n_{i=0}{\log p(\jmb{y}_i | \jmb{y}_{i-1}, \jmb{z}_i; \jmb{\theta}) + \log p(\jmb{z}_i; \jmb{\theta})}]}_{\text{Rate}}  \\
\end{equation}
where each $\jmb{x}_i, \jmb{\hat{x}}_i, \jmb{y}_i, \jmb{z}_i$ is a full/reconstructed video frame and code/hyperprior code respectively. The first term describes the reconstruction quality of the decoded video frames, and the second term measures the bitrate as approximated by our conditional entropy model. Each $\jmb{y}_i, \jmb{\hat{x}}_i$ is produced via our image encoder/decoder, while our conditional entropy model captures the dependence of $\jmb{y}_i$ on $\jmb{y}_{i-1}, \jmb{z}_i$. We can additionally clamp the rate term to 
$$max(\mathbb{E}_{x \sim p_{data}}[\sum^n_{i=0}{\log p(\jmb{y}_i | \jmb{y}_{i-1}, \jmb{z}_i; \jmb{\theta}) + \log p(\jmb{z}_i; \jmb{\theta})}], R_a)$$
 to enforce a target bitrate $R_a$.  


\section{Internal Learning of the Frame Code} \label{sec:test_time}

We additionally propose an internal learning extension of our base model, which leverages every frame of a test video sequence as its own example for which we can learn a better encoding, helping to provide more gains in rate-distortion performance with our entropy-focused approach.  

The goal of a compression algorithm is to find codes that can later be \textit{decoded} according to a codebook that does not change during encoding.
 This is also intuitively why we can not overfit our entire compression architecture to a single frame in a video sequence; this would imply that every video frame would require a separate decoder to decode. However, we make the observation that in our models, the trained decoder/hyperprior decoder represent our codebook; hence as long as the decoder and hyperprior decoder parameters remain fixed, we can actually optimize the encoder/hyperprior parameters or the latent codes themselves, $\jmb{y}_i$ and $\jmb{z}_i$, for every frame during inference. In practice we do the latter to reduce the number of parameters to optimize. 
 
One benefit of internal learning in our video compression setting is similar to that suggested by Campos et al. \cite{campos_cao}: the test distribution during inference is oftentimes different than the training distribution, and this is especially true in video settings, where the test distribution may have different artifacts, framerate, etc. Our base conditional entropy model may predict a higher entropy for test videos due to distributional shift. Hence, internal learning might help account for the shortcomings of out-of-distribution prediction by the encoder/hyperprior encoder. 
 
The second benefit is unique to our video compression setting: we can optimize each frame code to reduce the joint entropy in a way that the base approach cannot. In the base approach, there is a restriction of assuming that $\jmb{y}_i$ is produced by an independent single-image compression model without accounting for past frames as input. Yet there exist configurations of $\jmb{y}_i$ with the same reconstruction quality that are more easily predictive from $\jmb{y}_{i-1}$ in our entropy model $p(\jmb{y}_i | \jmb{y}_{i-1}, \jmb{z}_i)$. Performing internal learning allows us to more effectively search for a more optimal configuration of the frame code and hyperprior code $\jmb{z}^*_i$, $\jmb{y}^*_i$ such that $\jmb{y}^*_i$ can be more easily predicted by $\jmb{y}^*_{i-1}, \jmb{z}^*_i$ in the entropy model.
As a result, internal learning helps open up a wider search space of frame codes that can potentially have a lower joint entropy. 

To perform internal learning during inference, we optimize against a similar rate-distortion loss as in Eq. \ref{eq:rate_distortion_loss}:

\begin{equation} \label{eq:test_time_loss}
\begin{split} 
L_{\text{internal}}(\jmb{x}) = \sum^n_{i=0}{\ell(\jmb{x}_i, \jmb{\hat{x}}_i)}+
\lambda \left[\sum^n_{i=0}{\log p(\jmb{y}_i | \jmb{y}_{i-1}, \jmb{z}_i; \jmb{\theta}) + \log p(\jmb{z}_i; \jmb{\theta}) } \right] 
\end{split}
\end{equation}
where $\jmb{x}$ denotes the test video sequence that we optimize over, and $\ell$ represents the reconstruction loss function. We first initialize $\jmb{y}_i$ and $\jmb{z}_i$ as the output from the trained encoder/hyperprior encoder. Then we backpropagate gradients from (Eq. \ref{eq:rate_distortion_loss}) to $\jmb{y}_i$ and $\jmb{z}_i$ for a set number of steps, while keeping all decoder parameters fixed.
We can additionally customize $\lambda$ in Eq. (\ref{eq:test_time_loss}) depending on whether we want to tune more for bitrate or reconstruction. If the newly optimized codes are denoted as $\jmb{y}^*_i$ and $\jmb{z}^*_i$, then we simply store $\jmb{y}^*_i$ and $\jmb{z}^*_i$ during encoding and discard the original $\jmb{y}_i$ and $\jmb{z}_i$. 
We do note that internal learning during inference prevents the ability to perform parallel frame encoding, since $\jmb{y}^*_i, \jmb{z}^*_i$ now depend on $\jmb{y}^*_{i-1}$ as an output of internal learning rather than the image encoder; the gradient steps also increase the encoding runtime per frame. However, after $\jmb{z}_i$, $\jmb{y}_i$ are optimized, they are fixed during decoding, and hence decoding runtime does not increase. We will analyze the tradeoff of increased computation vs. reduced bitrates in the next section.

%% file: experiments.tex
\section{Experiments}

We present a detailed quantitative and qualitative analysis of our video compression approach on two datasets, varying factors such as frame-rate and video codec quality. 

\subsection{Datasets, Metrics, and Video Codecs} 
\paragraph{Kinetics, CDVL, and UVG, and others: }

We train on the Kinetics dataset \cite{carreira_kinetics}. Then, we benchmark our method against standard video test sets which are commonly used for evaluating video compression algorithms. Specifically, we run evaluations on video sequences from the Consumer Digital Video Library (CDVL) \cite{cdvl}  as well as the Ultra Video Group (UVG) \cite{uvg}. UVG consists of 7 video sequences of 3900 frames total, each $1920 \times 1080$ and 120fps. Our CDVL dataset consists of 78 video sequences, each $640 \times 480$ and either 30fps or 60fps. The videos span a wide range of natural image settings as well as motion. To demonstrate further analysis of our approach, we benchmark on video sequences from MCL-JVC \cite{mcl_jvc} and Video Trace Library (VTL) \cite{vtl}, which are shown in supplementary material. 
\vspace{-1mm}
\paragraph{NorthAmerica: } 
We collect a video dataset by driving our self-driving fleet in several North American cities and collecting monocular, frontal camera data. The framerate is 10 Hz. Our training set consists of 1160 video sequences of 300 frames each, and our test set consists of 68 video sequences of 300 frames each. All frames are 1920 $\times$ 1200 in resolution, and we train on 240 $\times$150 crops. We focus both on full street driving sequences as well as only on sequences where ego-vehicle is moving (so no red-lights or stop signs).

\vspace{-1mm}



\paragraph{Metrics:}
We measure runtime in milliseconds on a per-frame basis for both encoding and decoding. Moreover we plot the rate-distortion curve for multi-scale structural similarity (MS-SSIM) \cite{wang_msssim}, which is a commonly-used perceptual metric that captures the overall structural similarity in the reconstruction. We report the MS-SSIM curve at log-scale similar to \cite{balle_varhyperprior}, where log-scale is defined as $-10\log_{10}(1-\text{MS-SSIM})$. Additionally, we report some curves using PSNR:  $-10\log_{10}(\text{MSE})$, where MSE is mean-squared error, and hence measuring the absolute error in the reconstructed image. 

Note from Sec. \ref{sec:rate_distortion_loss} that all our base models are trained/optimized with mean-squared error (MSE), and we show that our base models are robust in both MS-SSIM and PSNR.  However with internal learning, we demonstrate the flexibility of tuning to different metrics during test-time, so we optimize reconstruction loss towards MS-SSIM and MSE separately (see Sec. \ref{sec:test_time}). 



\vspace{-1mm}

\begin{figure}
	\centering
	\includegraphics[width=0.45\linewidth]{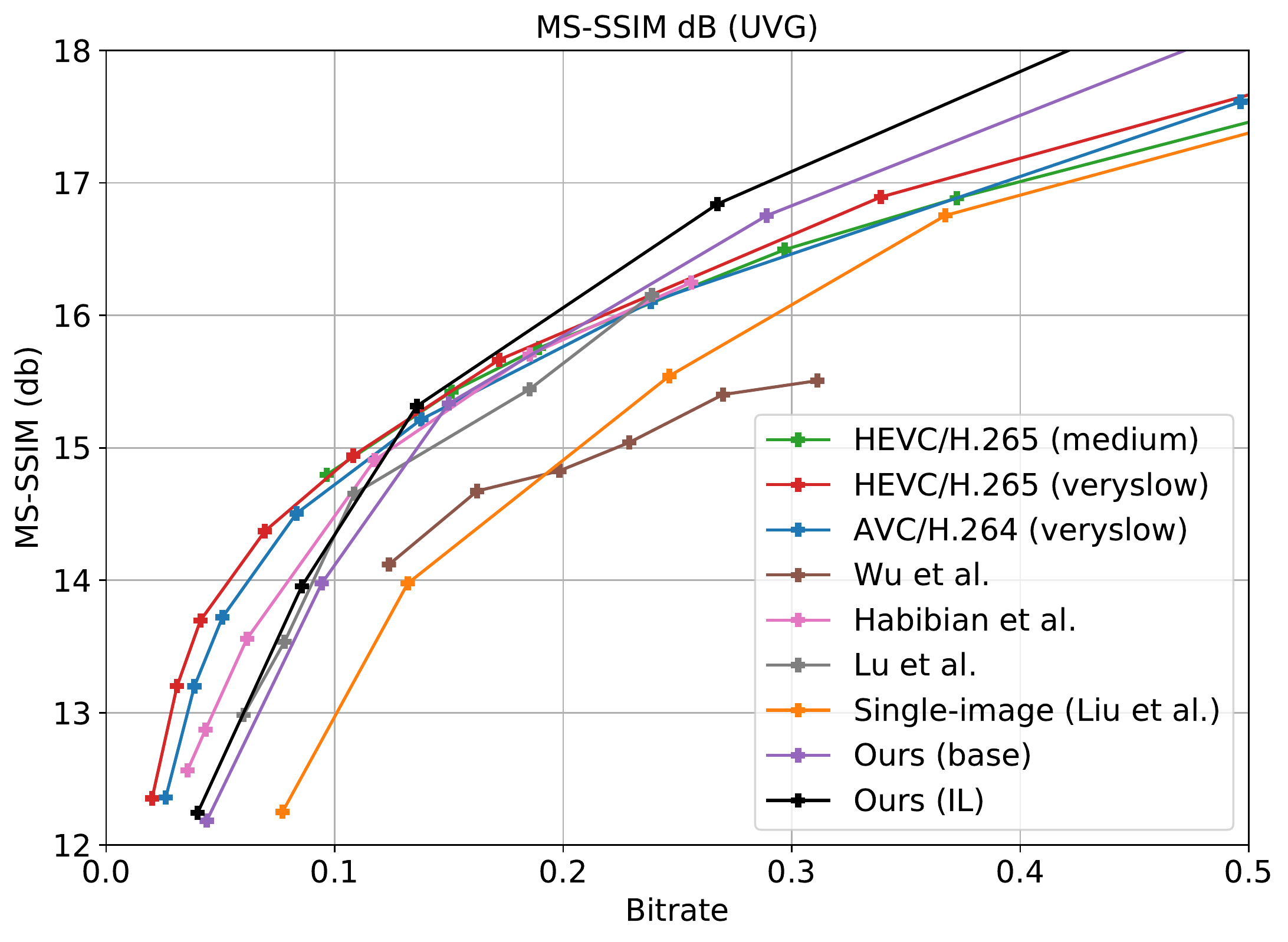}
	\includegraphics[width=0.45\linewidth]{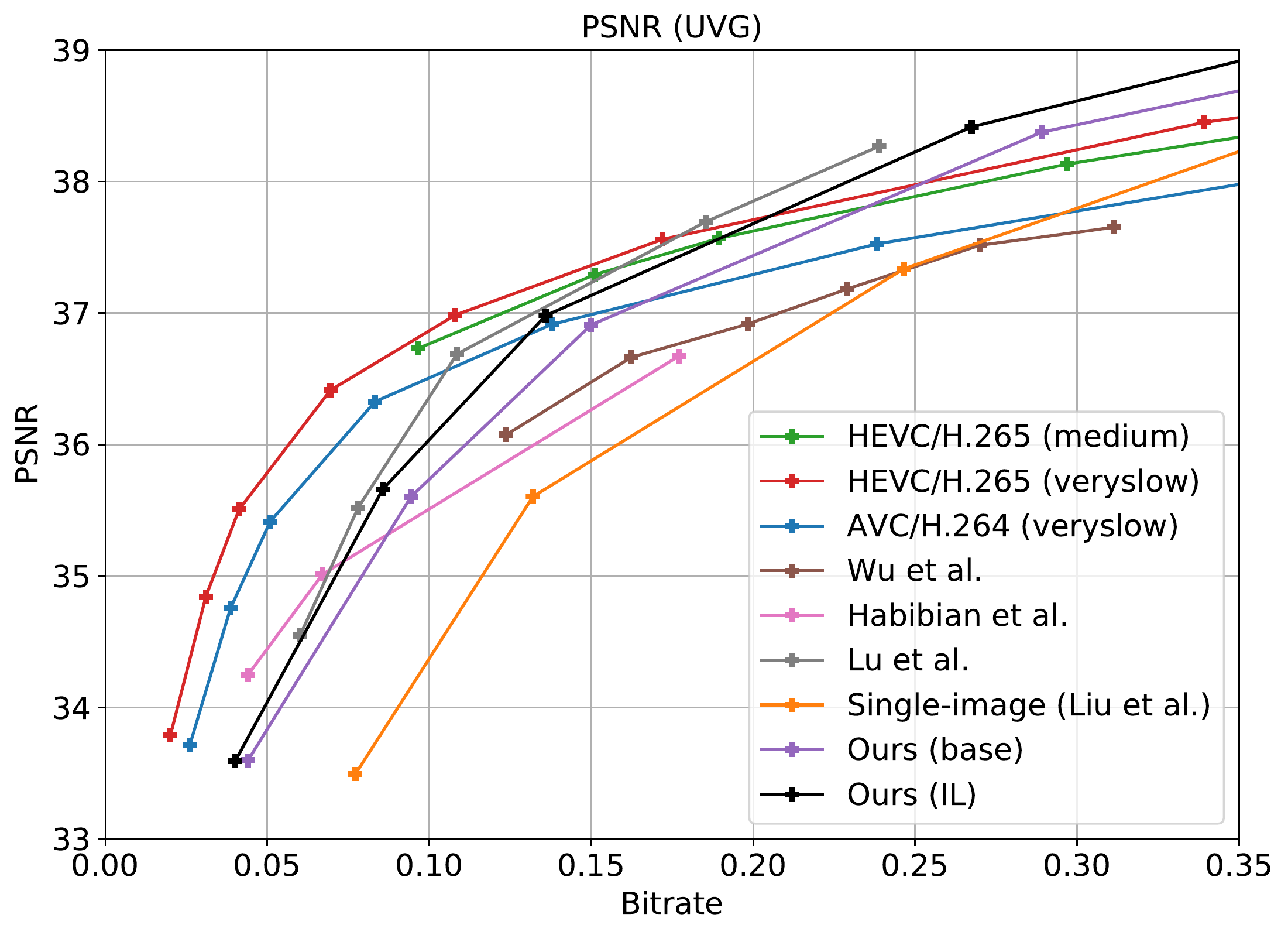}
	\vspace{-2mm}
	\caption{Rate-distortion plot of our model against competing deep compression works \cite{wu_vidinterpolation,habibian_rdauto,lu_dvc}. Results are on full 1920 $\times$ 1080 UVG video. 
	}
	\label{fig:comp_uvg_figure}
	\vspace{-4mm}
\end{figure}

\paragraph{Video Codecs and Baselines:} We benchmark with both libx265 (HEVC/H.265) and libx264 (AVC/H.264). To the best of our knowledge all prior works on learned video compression \cite{habibian_rdauto,rippel_learnedvidcomp,wu_vidinterpolation,djelouah_interframe,lu_dvc} have artificially restricted codec performance either by using a faster setting or by imposing additional limitations on the codecs (such as removing B-frames). In contrast, we benchmark both H.265 and H.264 on the \textit{veryslow} setting in ffmpeg in order to maximize the performance of these codecs. For the sake of illustration (and also to have a consistent comparison in Fig. \ref{fig:comp_uvg_figure} with other authors) we also plot H.265 with the \textit{medium} preset for benchmarking. We also incorporate corresponding numbers from the learned compression methods of \cite{wu_vidinterpolation}, \cite{lu_dvc}, \cite{habibian_rdauto}. Finally, we add our single-image compression model, inspired by \cite{liu_nonlocalcomp}, as a baseline.

In addition, we remove Group of Picture (GoP) restrictions when running H.265/H.264, such that the maximum GoP size is equivalent to the total number of frames of each video sequence. We note that neither our base approach nor internal learning require an explicit notion of GoP size: in the base approach, every frame code is produced independently with an image encoder, and with internal learning we optimize every frame sequentially.  


\newcommand{\jcaptiond}[1]{\put(0,2){\sffamily \scriptsize \colorbox{gray}{\color{white} #1}}}
\newcommand{\jcaptiondd}[1]{\put(0,3){\sffamily \scriptsize \colorbox{gray}{\color{white} #1}}}
\newcommand{\jcaptione}[1]{\put(0,45){\sffamily \tiny \colorbox{gray}{\color{white} #1}}}
\newcommand{\recona}[3]{
	\begin{overpic}[width=\imw]{#1}
		\put(13,19){ \color{green} \framebox(6, 8){}} 
		\put(64,0){\setlength{\fboxsep}{0pt}\color{green}\fbox{\includegraphics[viewport=294 364 411 518, clip, height=50pt]{#1}}}
		\jcaptiond{#2}
		\jcaptione{#3}
\end{overpic}}
\newcommand{\reconb}[3]{
	\begin{overpic}[width=\imw]{#1}
		\put(13,19){ \color{green} \framebox(6, 8){}} 
		\put(64,0){\setlength{\fboxsep}{0pt}\color{green}\fbox{\includegraphics[viewport=294 364 411 518, clip, height=50pt]{#1}}}
		\jcaptiondd{#2}
		\jcaptione{#3}
\end{overpic}}
\newcommand{\jcaptionf}[1]{\put(0,2){\sffamily \scriptsize \colorbox{gray}{\color{white} #1}}}
\newcommand{\jcaptiong}[1]{\put(0,59){\sffamily \tiny \colorbox{gray}{\color{white} #1}}}
\newcommand{\trecona}[3]{
	\begin{overpic}[width=\imw]{#1}
		\put(18,25){ \color{green} \framebox(8, 8){}} 
		\put(54,0){\setlength{\fboxsep}{0pt}\color{green}\fbox{\includegraphics[viewport=392 480 548 640, clip, height=50pt]{#1}}}
		\jcaptionf{#2}
		\jcaptiong{#3}
\end{overpic}}
\newcommand{\jcaptionff}[1]{\put(0,3){\sffamily \scriptsize \colorbox{gray}{\color{white} #1}}}
\newcommand{\treconb}[3]{
	\begin{overpic}[width=\imw]{#1}
		\put(18,25){ \color{green} \framebox(8, 8){}} 
		\put(54,0){\setlength{\fboxsep}{0pt}\color{green}\fbox{\includegraphics[viewport=392 480 548 640, clip, height=50pt]{#1}}}
		\jcaptionff{#2}
		\jcaptiong{#3}
\end{overpic}}

\newcommand{\jcaptionh}[1]{\put(0,2){\sffamily \scriptsize \colorbox{gray}{\color{white} #1}}}
\newcommand{\jcaptioni}[1]{\put(0,69){\sffamily \tiny \colorbox{gray}{\color{white} #1}}}
\newcommand{\crecona}[3]{
	\begin{overpic}[width=\imw]{#1}
		\put(49,13){ \color{green} \framebox(14,14){}} 
		\put(0,0){\setlength{\fboxsep}{0pt}\color{green}\fbox{\includegraphics[viewport=330 90 421 188, clip, height=60pt]{#1}}}
		\jcaptionh{#2}
		\jcaptioni{#3}
\end{overpic}}
\newcommand{\jcaptionhh}[1]{\put(0,3){\sffamily \scriptsize \colorbox{gray}{\color{white} #1}}}
\newcommand{\creconb}[3]{
	\begin{overpic}[width=\imw]{#1}
		\put(49,13){ \color{green} \framebox(14, 14){}} 
	\put(0,0){\setlength{\fboxsep}{0pt}\color{green}\fbox{\includegraphics[viewport=330 90 421 188, clip, height=60pt]{#1}}}
		\jcaptionhh{#2}
		\jcaptioni{#3}
\end{overpic}}

\begin{figure*} [!htb]
	\centering
	\def\imw{0.32\textwidth}
	\setlength{\tabcolsep}{1pt}
	\begin{tabular}{ccc}
		\recona{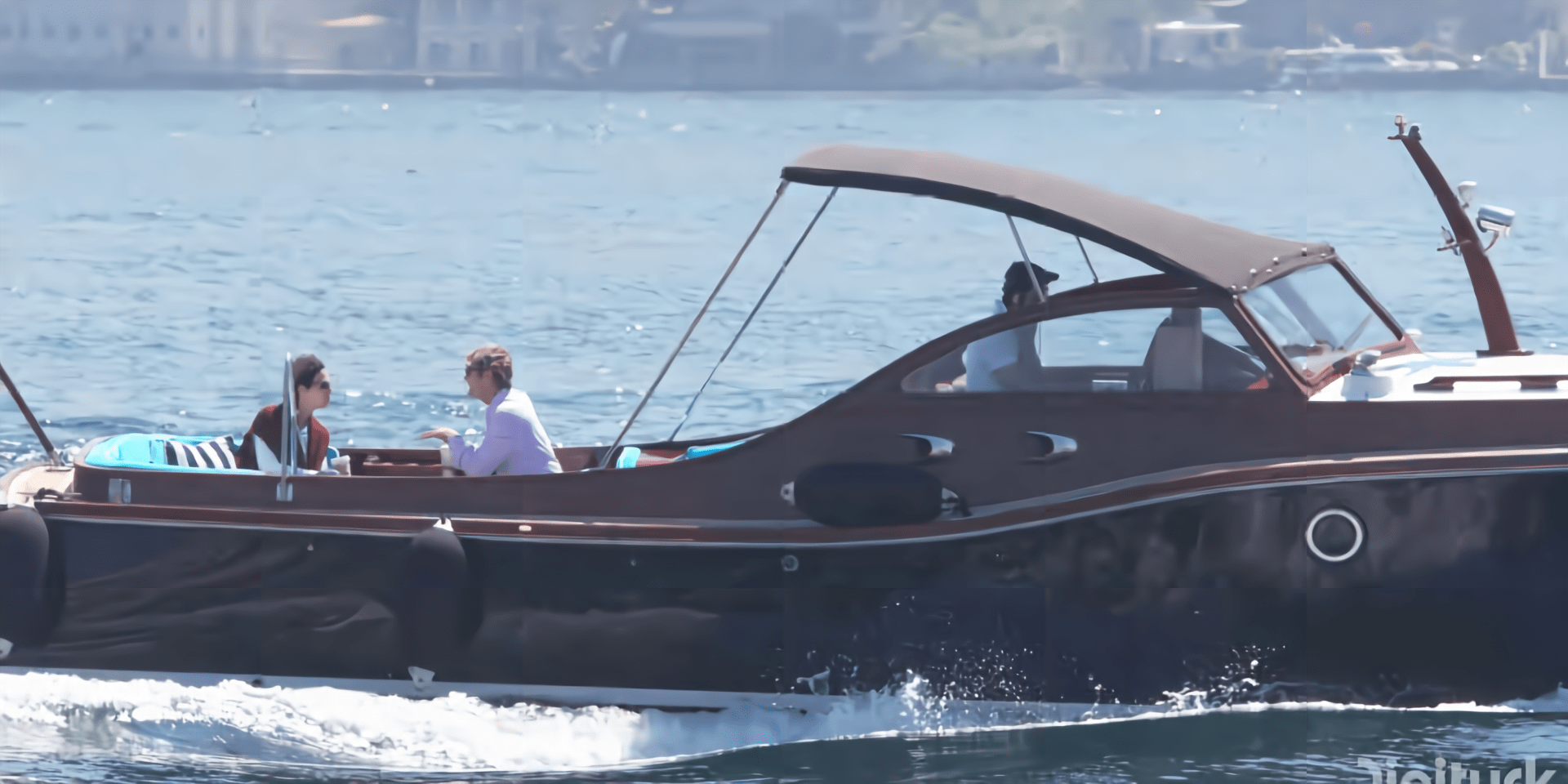}{\textbf{Ours}}{\textbf{[UVG] BPP: 0.076, MS-SSIM: 0.948}} &
		\reconb{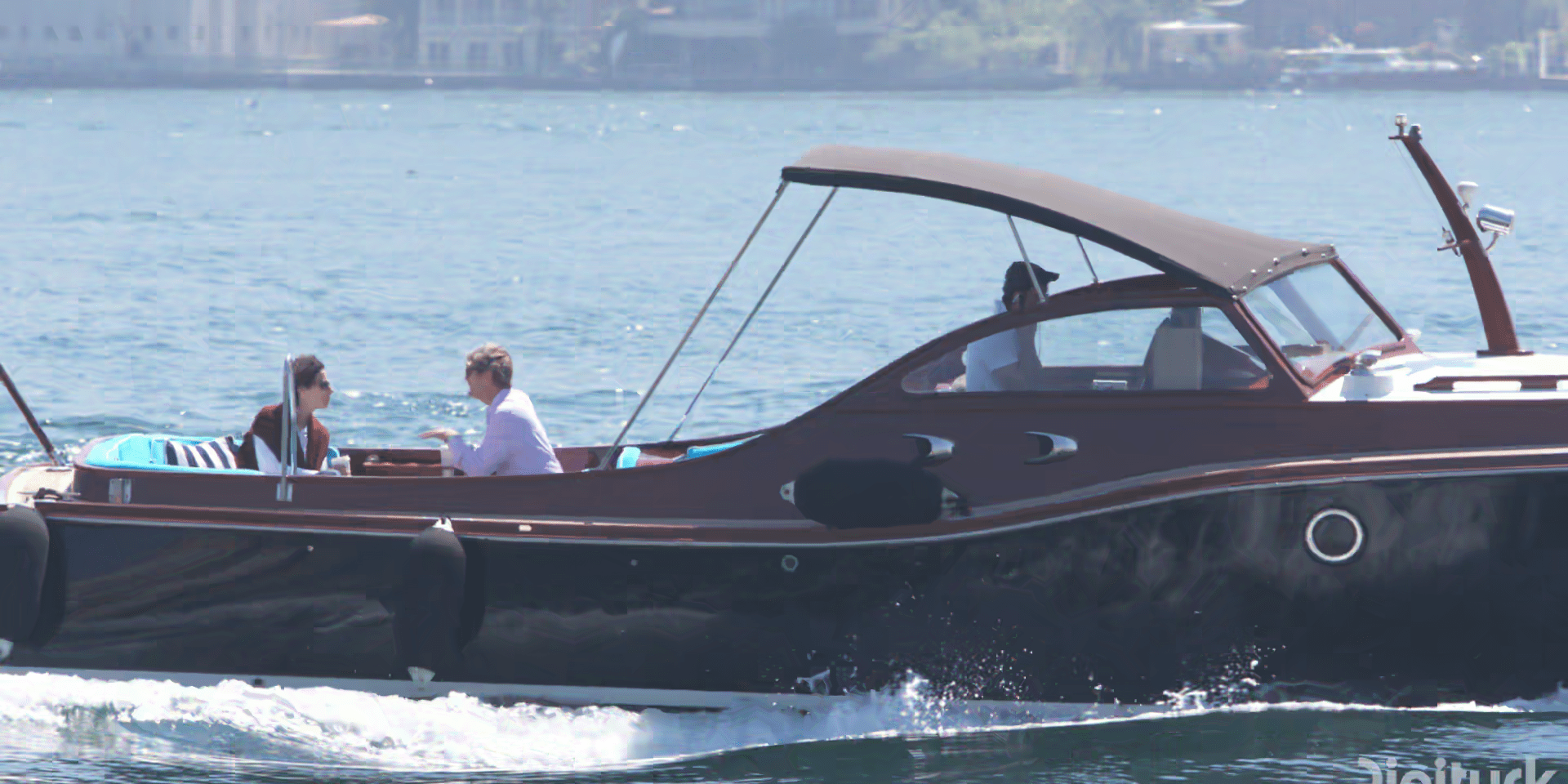}{H.265 (veryslow)}{[UVG] BPP: 0.121, MS-SSIM: 0.943} &
		\reconb{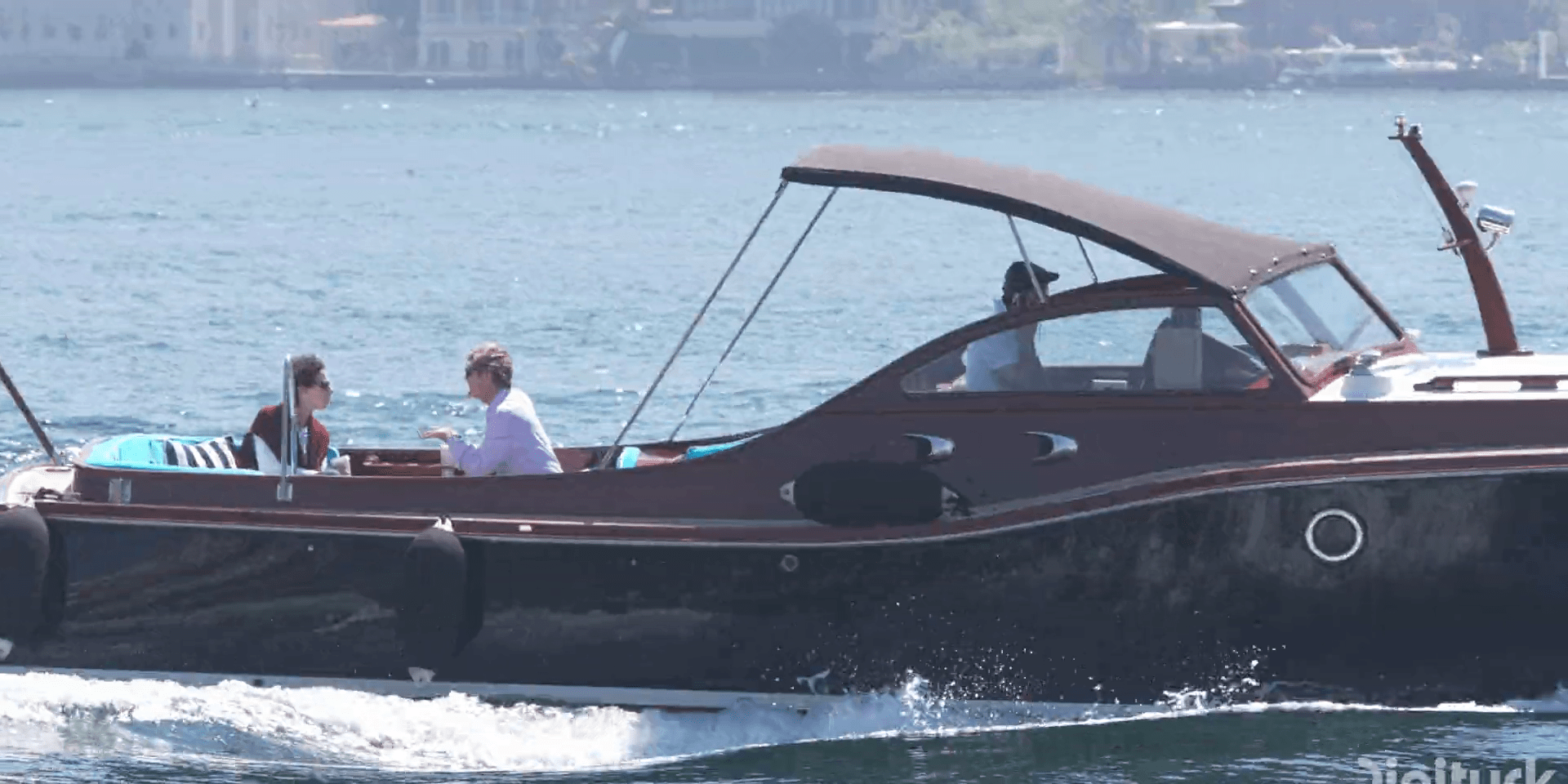}{H.264 (veryslow)}{[UVG] BPP: 0.082, MS-SSIM: 0.930}
		\\ [3pt]
		\trecona{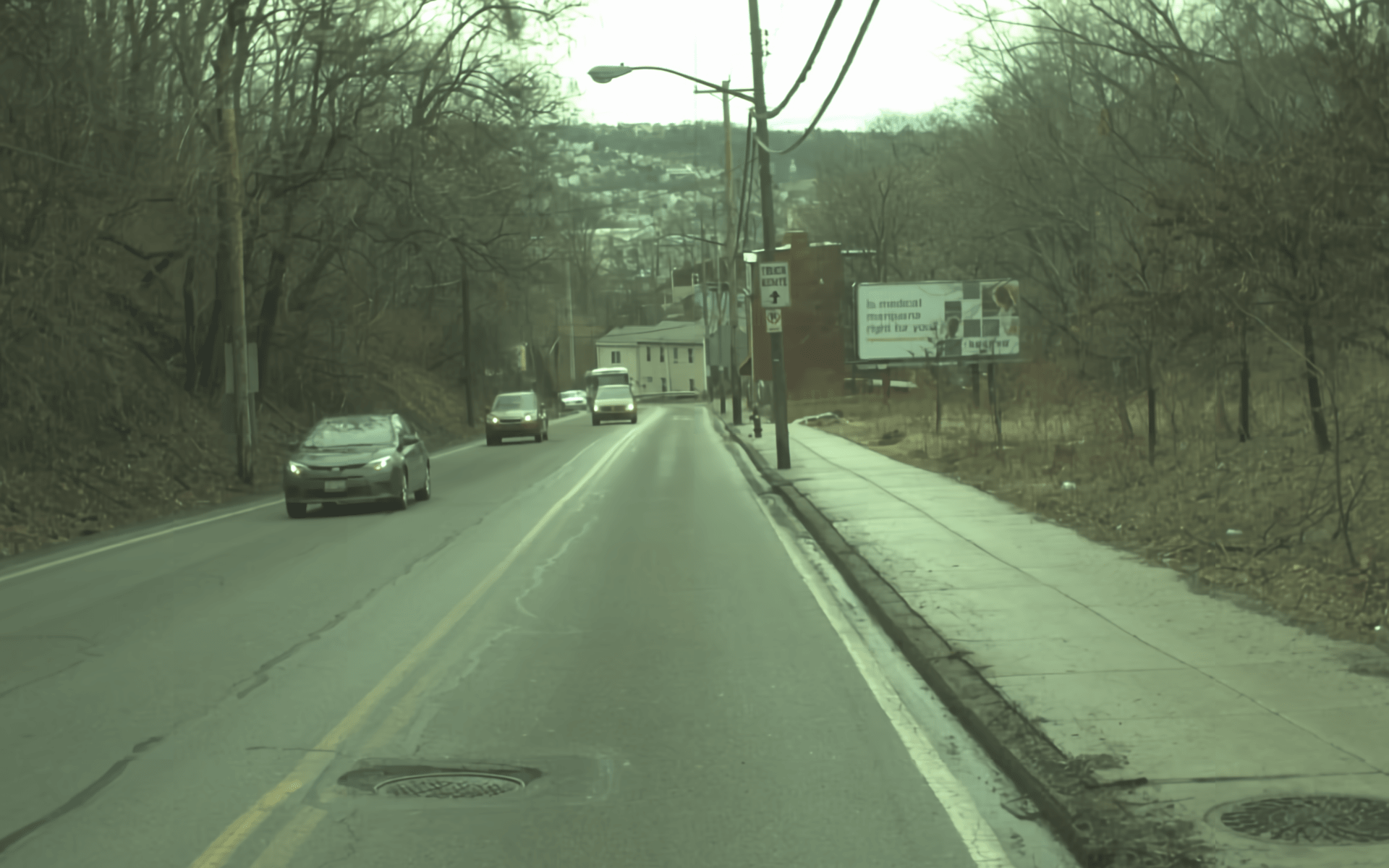}{\textbf{Ours}}{\textbf{[NorthAmerica] BPP: 0.087, MS-SSIM: 0.969}} &
		\treconb{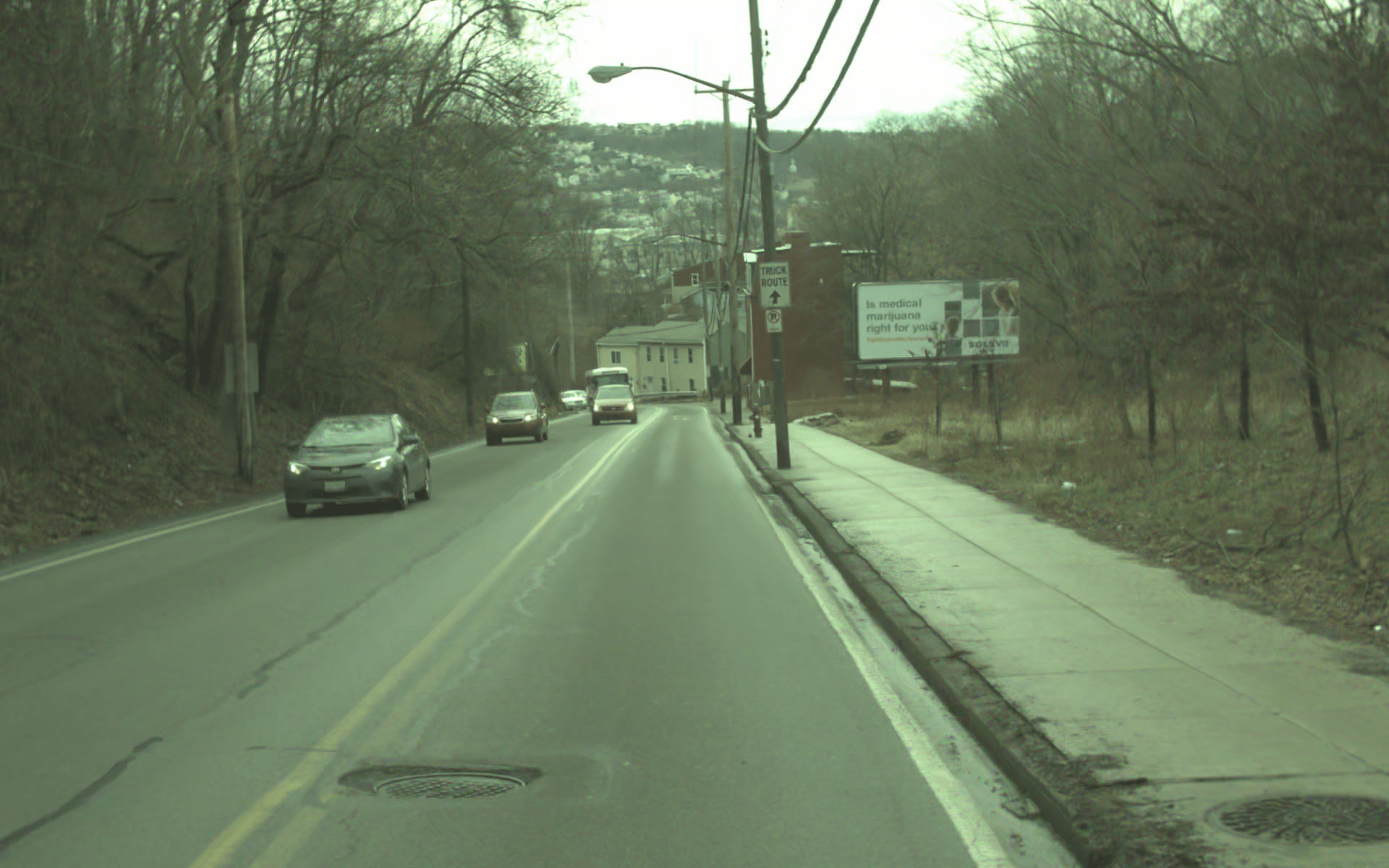}{H.265 (veryslow)}{[NorthAmerica] BPP: 0.107, MS-SSIM: 0.944} &
		\treconb{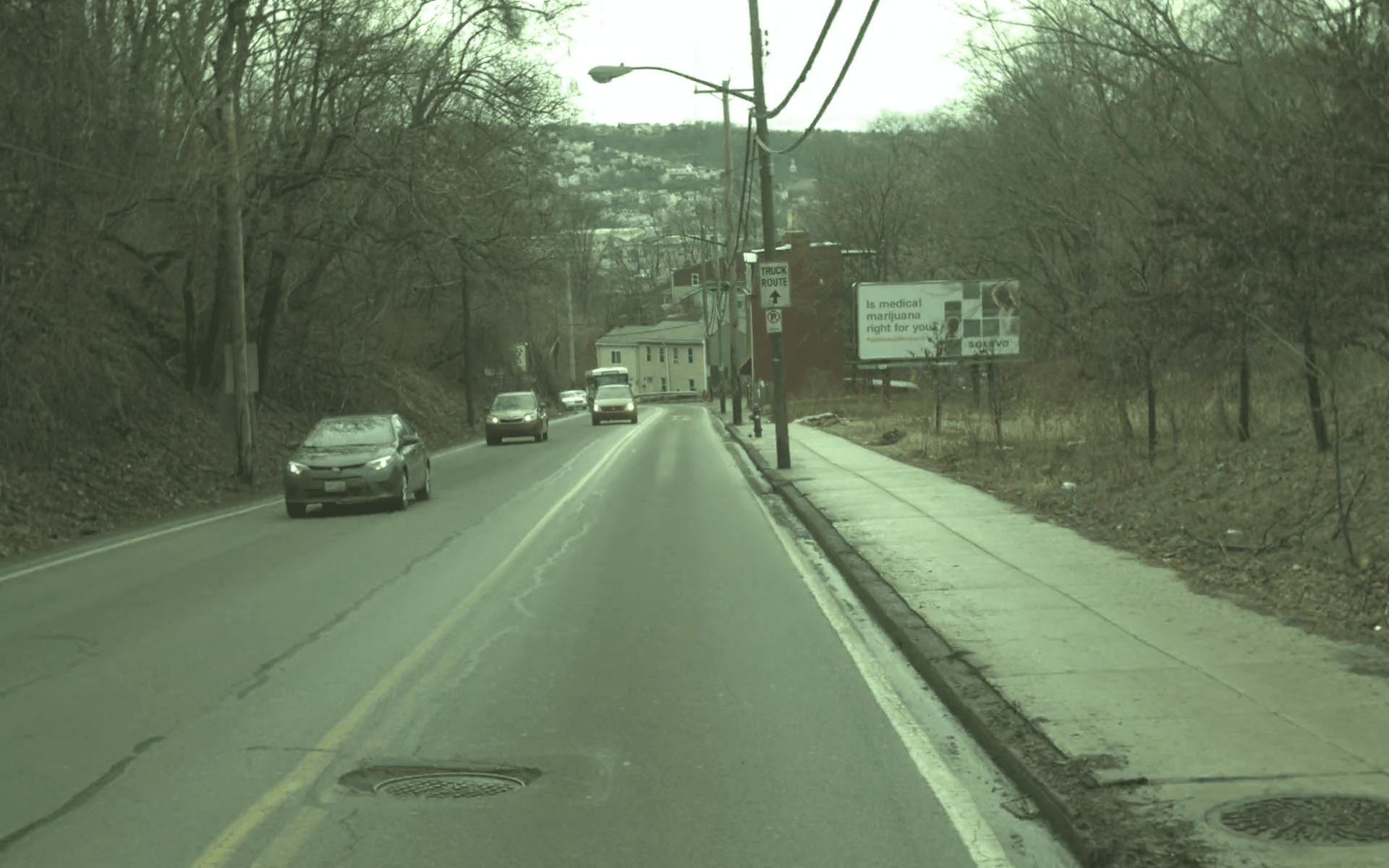}{H.264 (veryslow)}{[NorthAmerica] BPP: 0.097, MS-SSIM: 0.962}
		\\ [3pt]
		\crecona{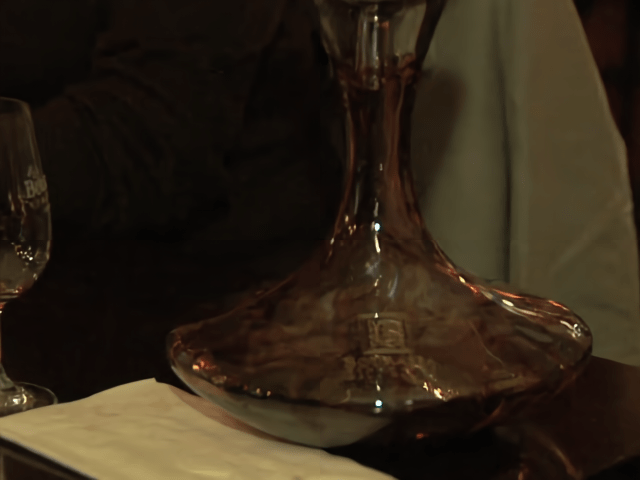}{\textbf{Ours}}{\textbf{[CDVL] BPP: 0.158, MS-SSIM: 0.969}} &
		\creconb{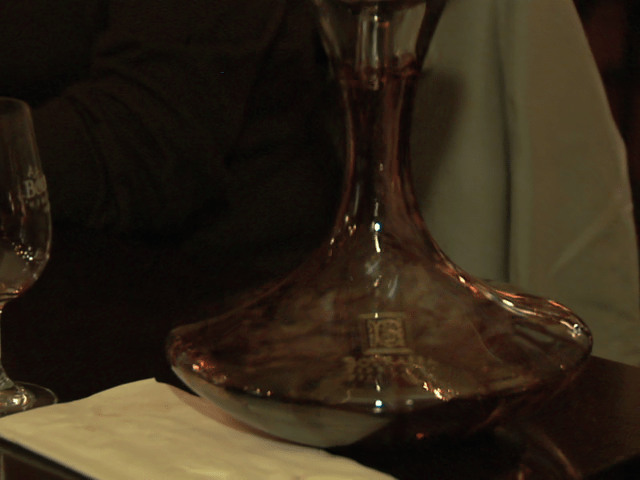}{H.265 (veryslow)}{[CDVL] BPP: 0.171, MS-SSIM: 0.967} &
		\creconb{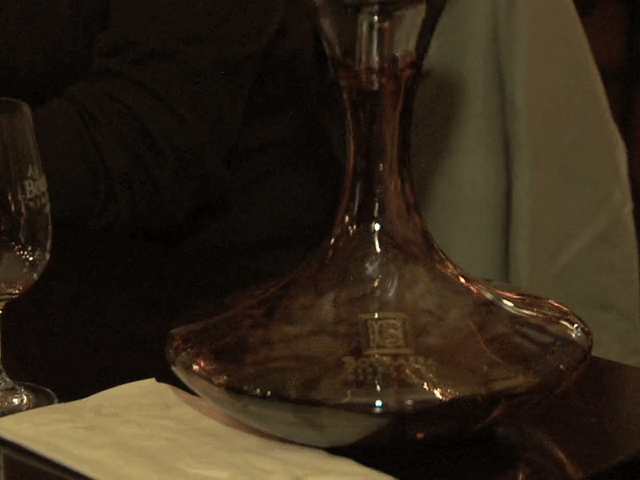}{H.264 (veryslow)}{[CDVL] BPP: 0.206, MS-SSIM: 0.960}
		\\ [-1pt]
		
	\end{tabular}
	\caption{Demonstration of our approach vs H.265 / H.264 on 10 Hz 1920 $\times$ 1200 NorthAmerica video, 12 Hz 1920 $\times$ 1080 UVG video, and 6 Hz $640 \times 480$ CDVL video. Even at lower bitrates, our approach demonstrates significant reductions in bitrate and distortion on lower framerate video.}
	\label{fig:qual_eval_all}
\end{figure*}


\vspace{-5mm}

\paragraph{Implementation Details:}
We use a learning rate of $7 \cdot 10^{-5}$ to $2 \cdot 10^{-4}$ for our models at different bitrates, and optimize parameters with Adam. We train with a batch size of 4 on two GPU's. For test/runtime evaluations, we use a single Intel Xeon E5-2687W CPU and a single 1080Ti GPU. For internal learning we run 10-12 steps of gradient descent per frame. Our range coding implementation is written in C++ interfacing with Python; during encoding/decoding we compute the codes and distributions on GPU, then pass the information over to our C++ implementation. 

%
%
%

\subsection{Runtime and Rate-distortion on UVG}
We showcase runtime vs. MS-SSIM plots of our method (both the base model and internal learning extension) against related deep compression works on UVG 1920 $\times$ 1080 video: Wu et al. \cite{wu_vidinterpolation}, Lu et al. \cite{lu_dvc}, and Habibian et al. \cite{habibian_rdauto}. \footnote{We don't show the results of Djelouah et al. \cite{djelouah_interframe} and Rippel et al. \cite{rippel_learnedvidcomp} because we were unable to get consistent MS-SSIM metrics on the UVG dataset. } Results are shown in Fig. \ref{fig:comp_runtime_figure}, and detail the frame encoding/decoding runtimes on GPU excluding the specific entropy coding implementation. \footnote{We thank the authors for providing us detailed runtime information.}


Overall our base approach is significantly faster than most deep compression works. During decoding, our base approach is \textit{orders of magnitude} faster than approaches that use an autoregressive entropy model (Habibian et al. \cite{habibian_rdauto}, Wu et al. \cite{wu_vidinterpolation}). We note that closest works in the GPU runtime and MS-SSIM is Lu et al., \cite{lu_dvc} who reported 666 ms for encoding and 556 ms for decoding. Nevertheless, our GPU-only pass is still faster (340ms for encoding and 191ms for decoding). 
Our entropy coding implementation has room for optimization; the C++ algorithm itself is fast (140 ms for range encoding, 139 ms for range decoding of a 1080p frame) though the {Python binding} interfacing brings the time up to 1.19/0.65 seconds for encoding and decoding. 


While the optional internal learning extension improves the rate-distortion trade-off in all the benchmarks, it brings overhead in encoding runtime. We note that our implementation of internal learning is unoptimized with the standard backward operator in PyTorch. However, it brings no overhead in decoding runtime, meaning our approach is still faster than all other approaches during decoding. 

In addition, we evaluate all the competing algorithms' performance and plot the rate-distortion curve on UVG test dataset, as shown in Fig. \ref{fig:comp_uvg_figure}. The results demonstrate that our approach is competitive or even outperforms existing approaches. 
Between bitrate ranges 0.1-0.3, which is where other deep baselines present their numbers, our base approach is as competitive as a motion-compensation approach \cite{lu_dvc} or one that uses autoregressive entropy models \cite{habibian_rdauto}. At higher bitrates, the base approach outperforms H.265 \textit{veryslow} in both MS-SSIM and PSNR. Internal learning further improves upon all bitrates by $\sim 10\%$.


\subsection{Rate-distortion on NorthAmerica}

\begin{figure*}
		\vspace{-1mm}
	\centering
	\includegraphics[height=0.22\linewidth]{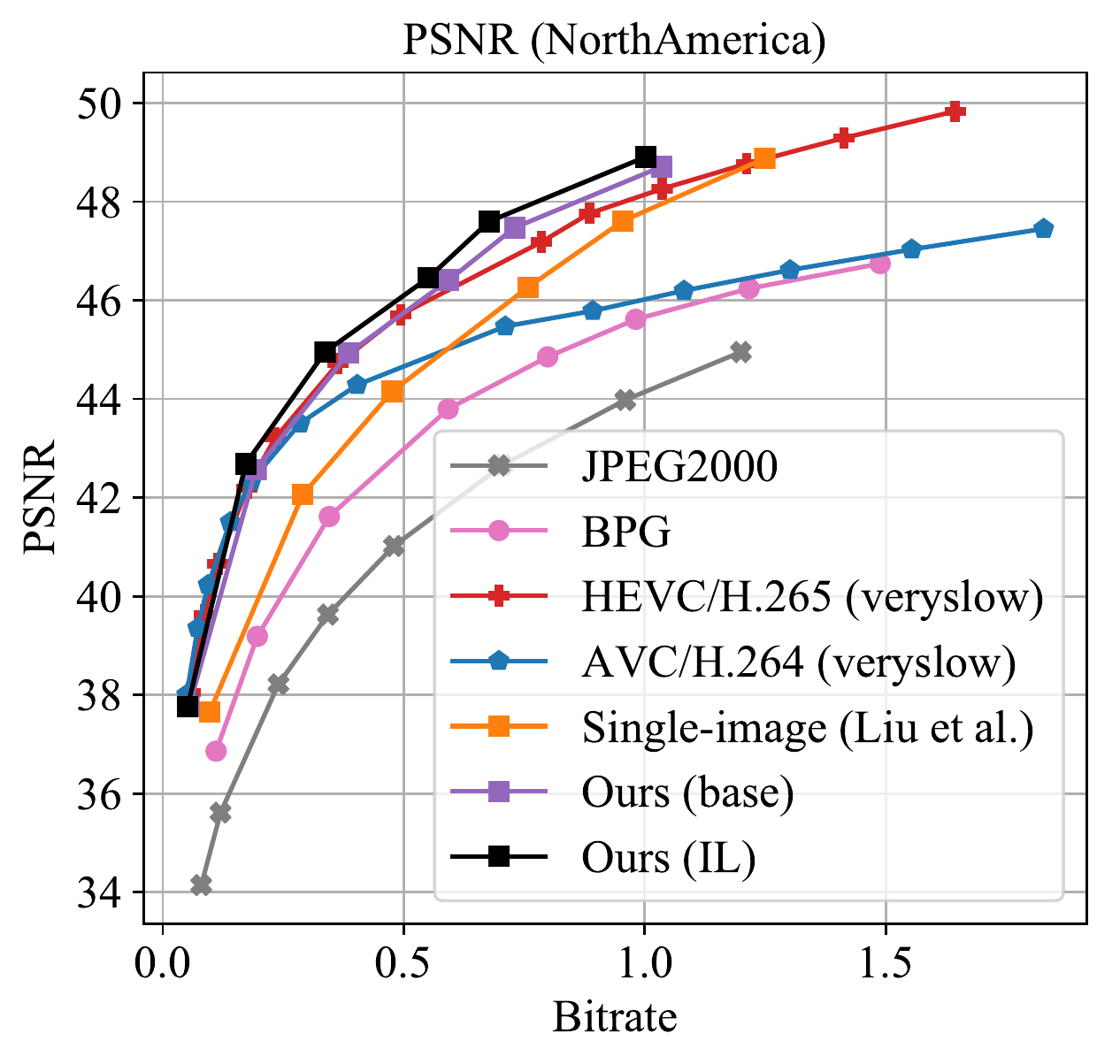}
	\includegraphics[height=0.22\linewidth]{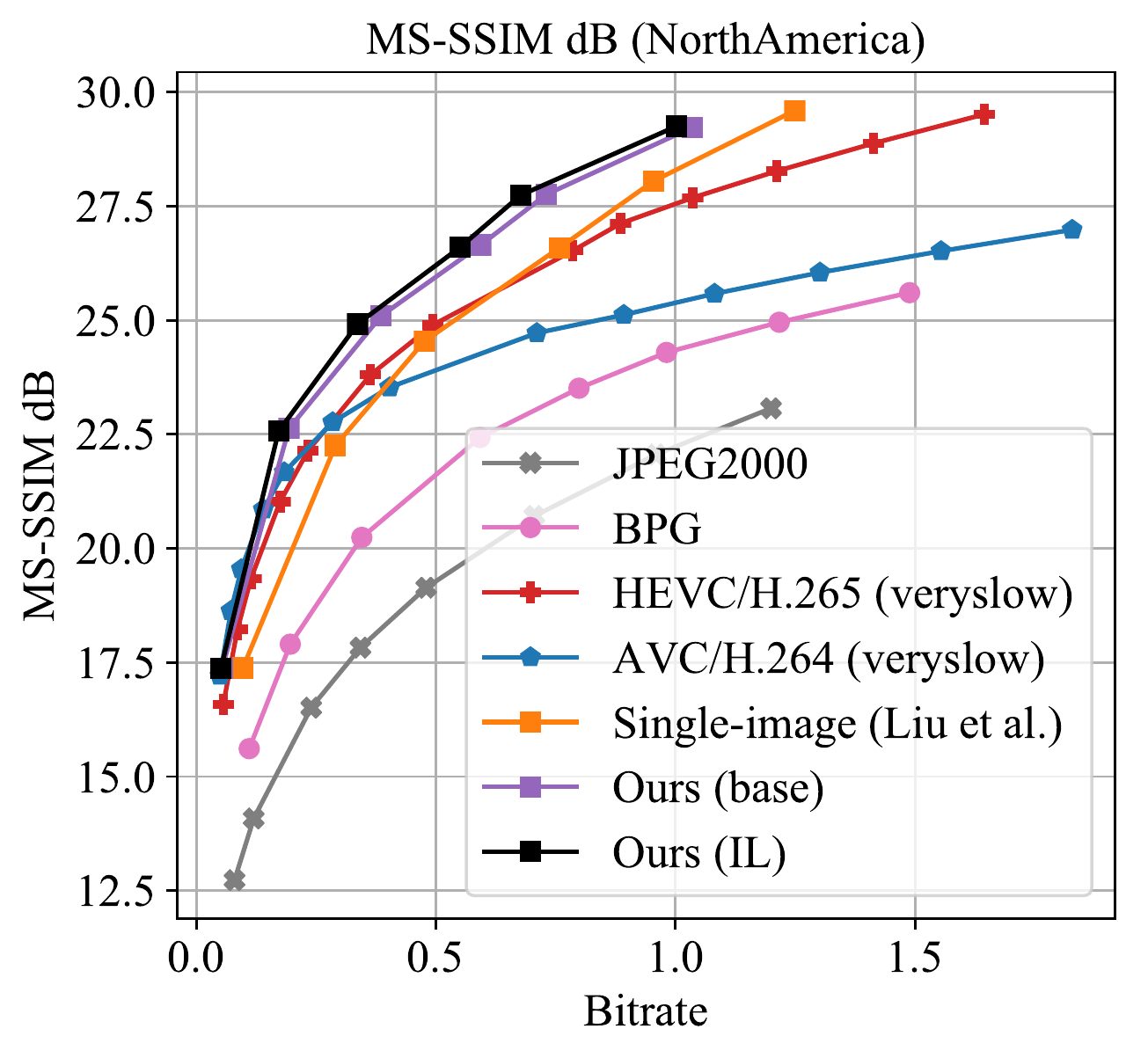} 
	\includegraphics[height=0.22\linewidth]{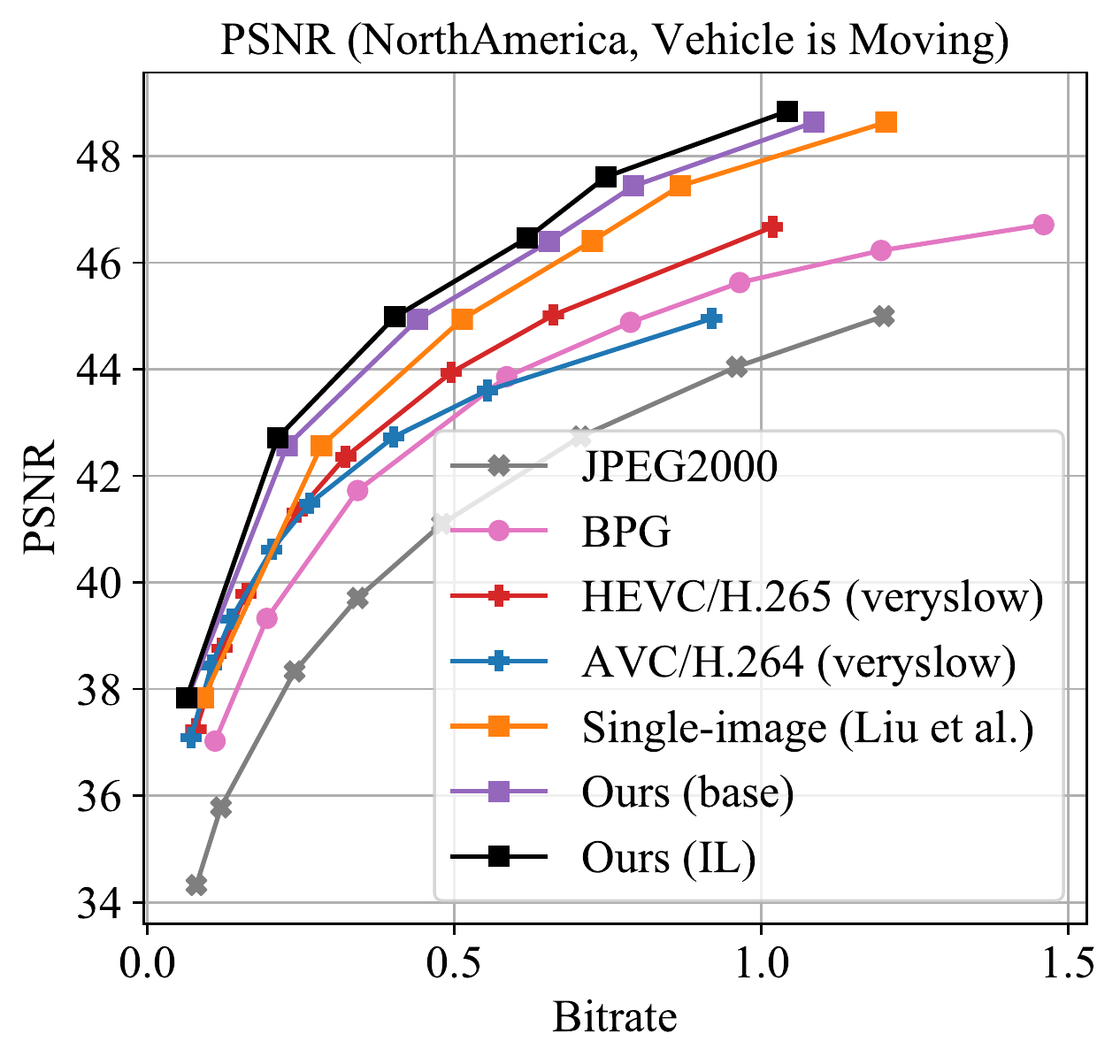}
	\includegraphics[height=0.22\linewidth]{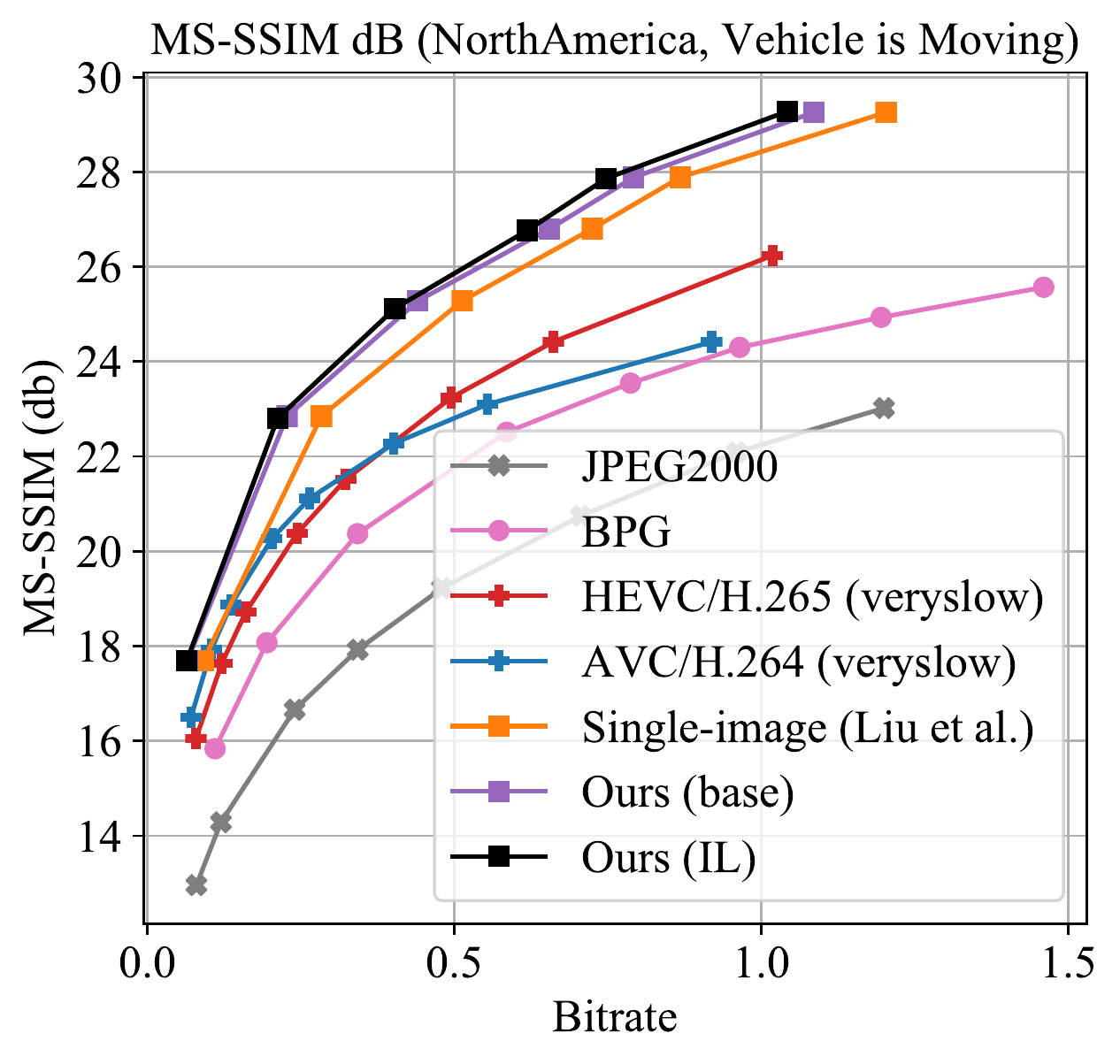}
	\vspace{-1mm}
	\caption{Plot of our approaches compared against compression baselines for NorthAmerica, both over the entire dataset as well as only when the ego-vehicle has positive velocity.}
	\label{fig:tor4d_results_figure}
	\vspace{-2mm}
\end{figure*}

We  show our conditional entropy model and internal learning extension on the NorthAmerica dataset, in Fig. \ref{fig:tor4d_results_figure}. The graph shows that even our single-image Liu model baseline \cite{liu_nonlocalcomp} outperforms H.265 on MS-SSIM at higher bitrates and approaches H.265 in PSNR. Our conditional entropy model demonstrates bitrate improvements of 20-50\% across bitrates, and internal learning demonstrates an additional ~10\% improvement. 

Fig. \ref{fig:tor4d_results_figure} also shows graphs in which we only analyze video sequences where the autonomous vehicle is in motion, which creates a fairly large gap in H.265 performance. In this setting, both our video compression algorithm as well as the single-image model outperform H.265 by a wide margin on almost all bitrates in MS-SSIM and at higher bitrates in PSNR.

\begin{figure*}
	\centering
	\includegraphics[height=0.23\linewidth]{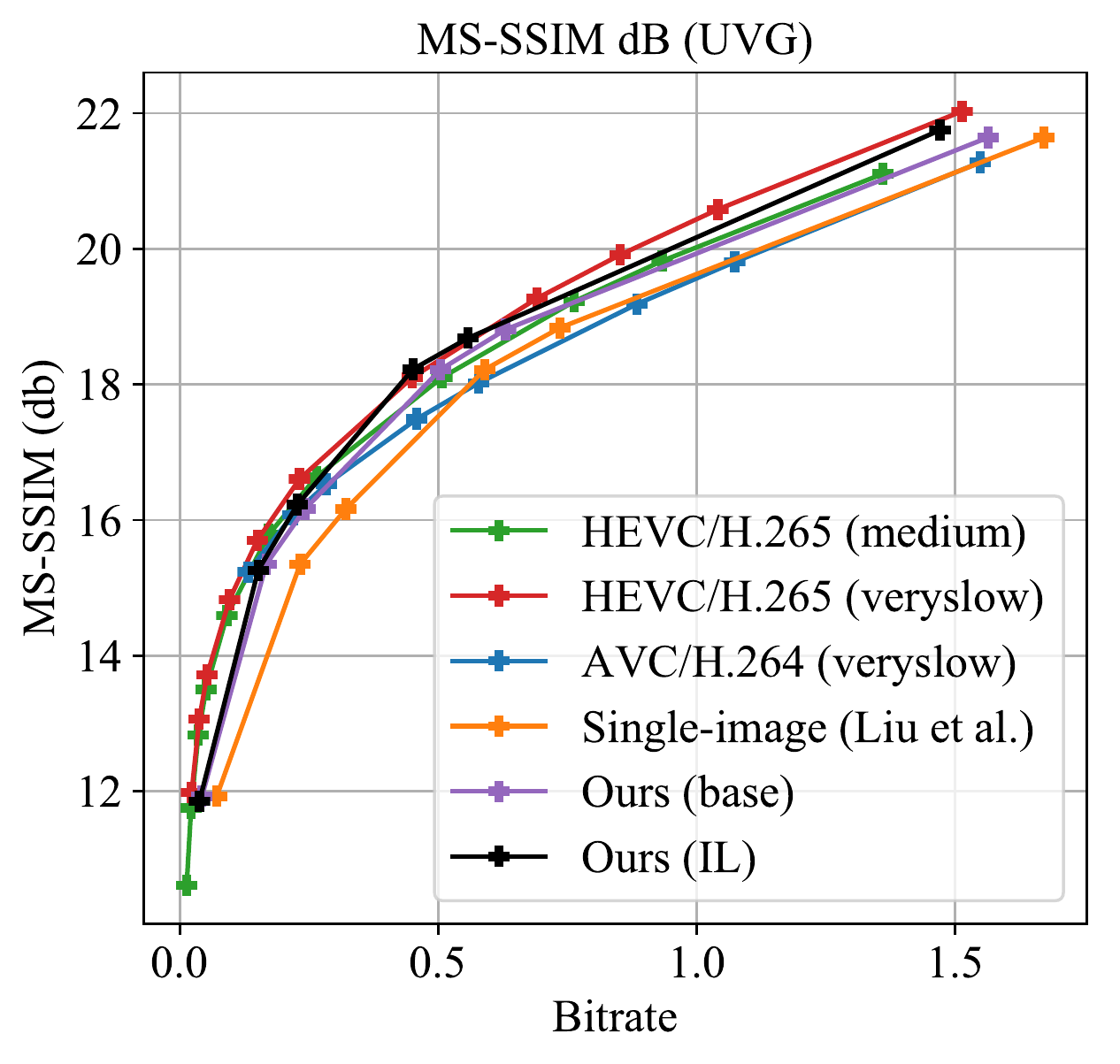}
	\includegraphics[height=0.23\linewidth]{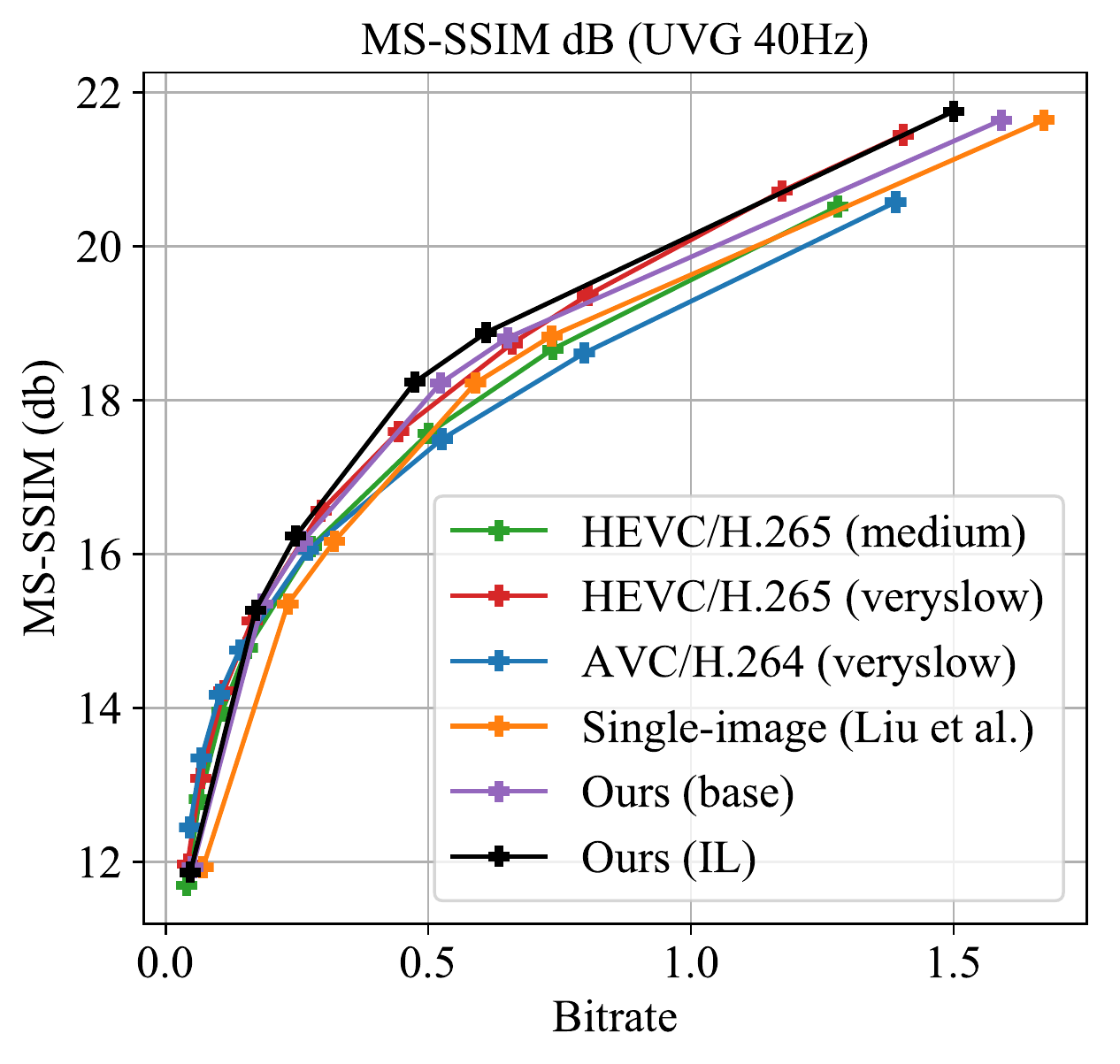}
	\includegraphics[height=0.23\linewidth]{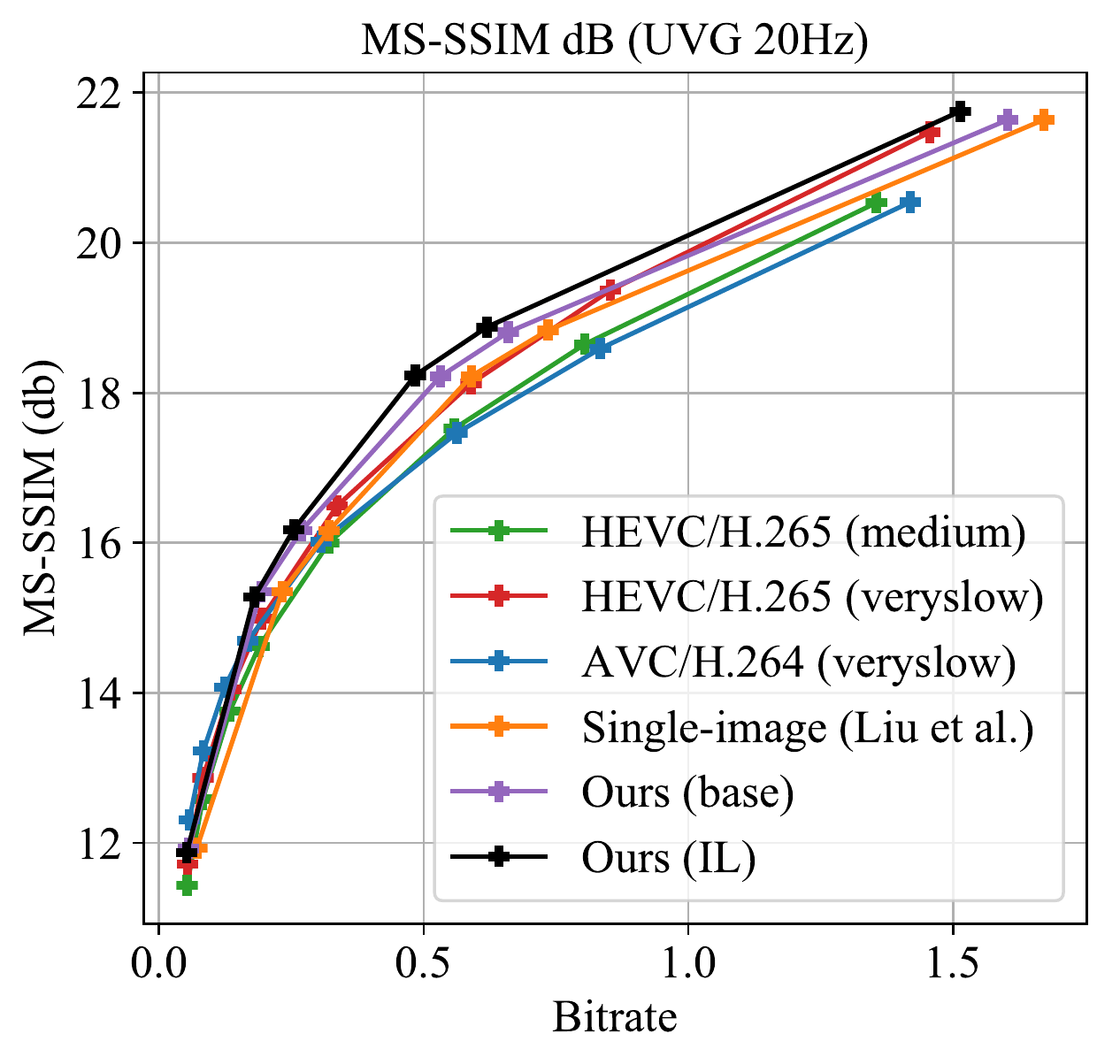} 
	\includegraphics[height=0.23\linewidth]{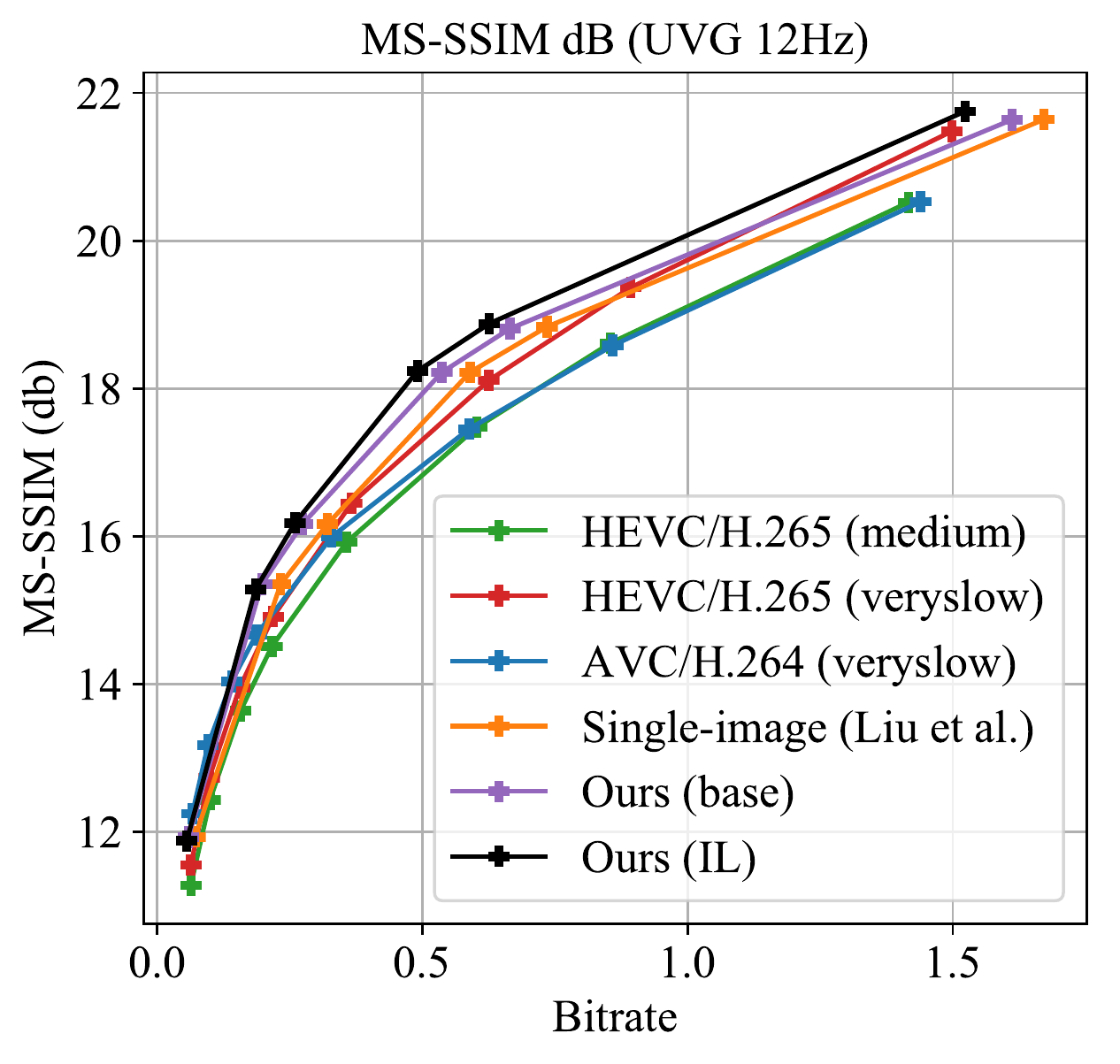} 
	\\
	\includegraphics[height=0.23\linewidth]{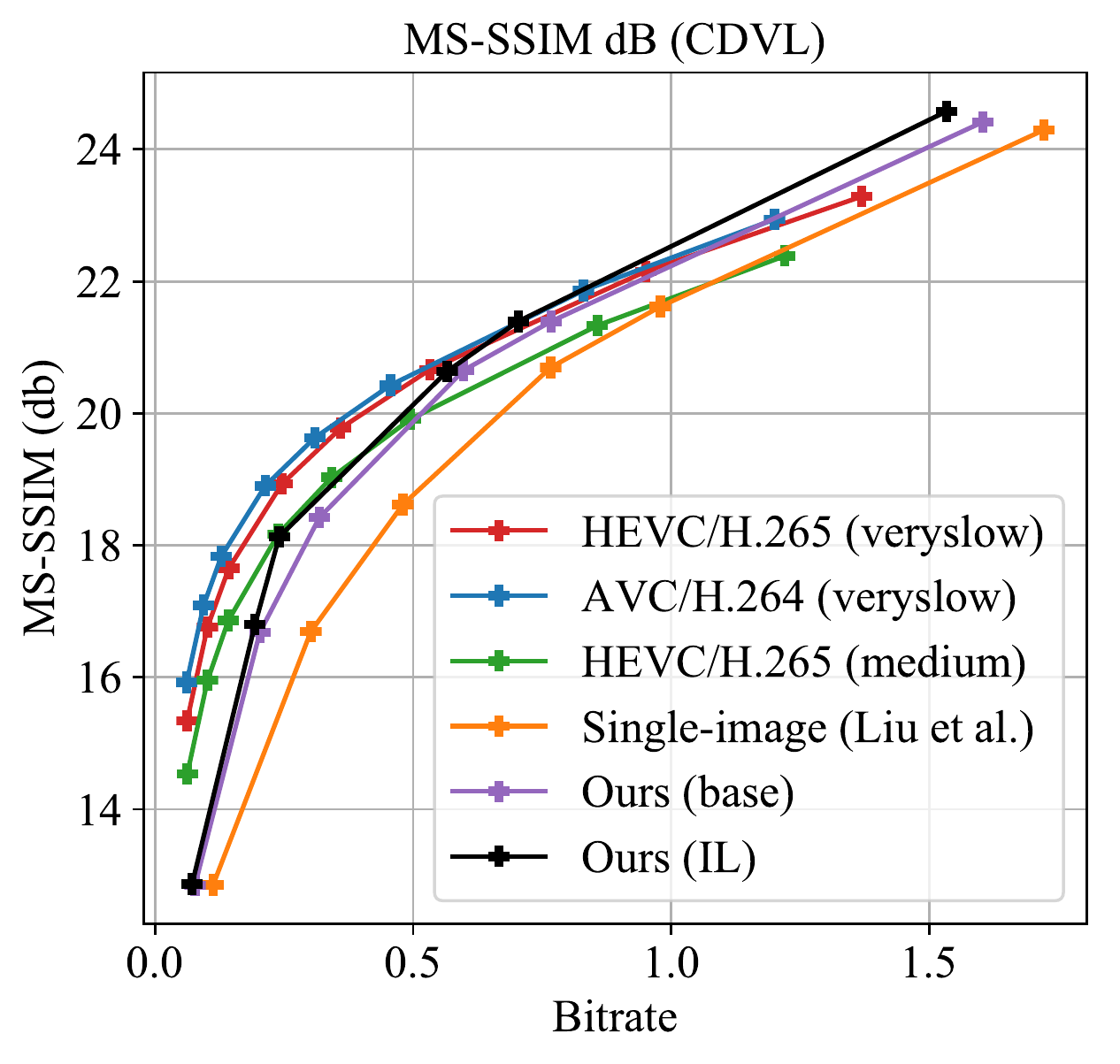}
	\includegraphics[height=0.23\linewidth]{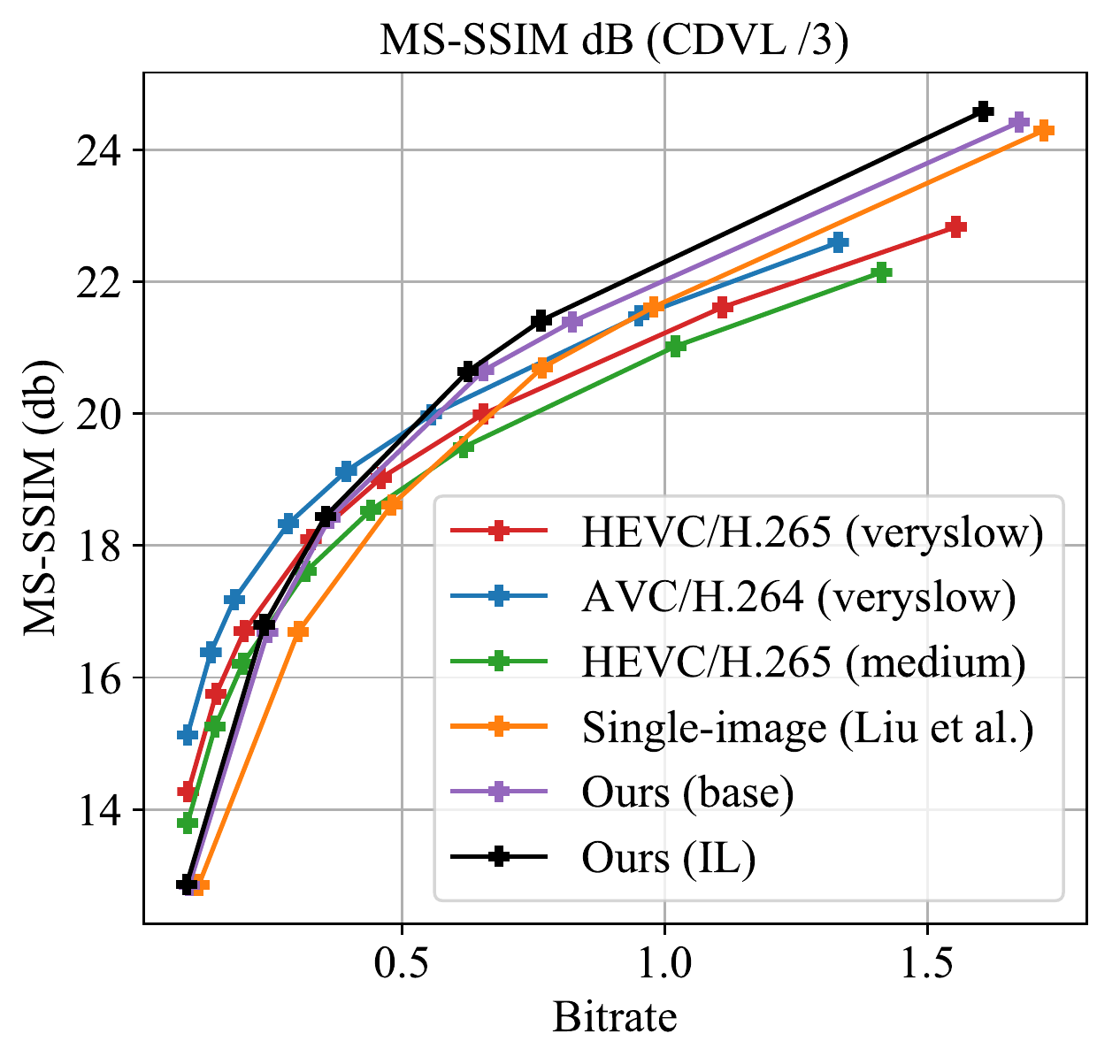}
	\includegraphics[height=0.23\linewidth]{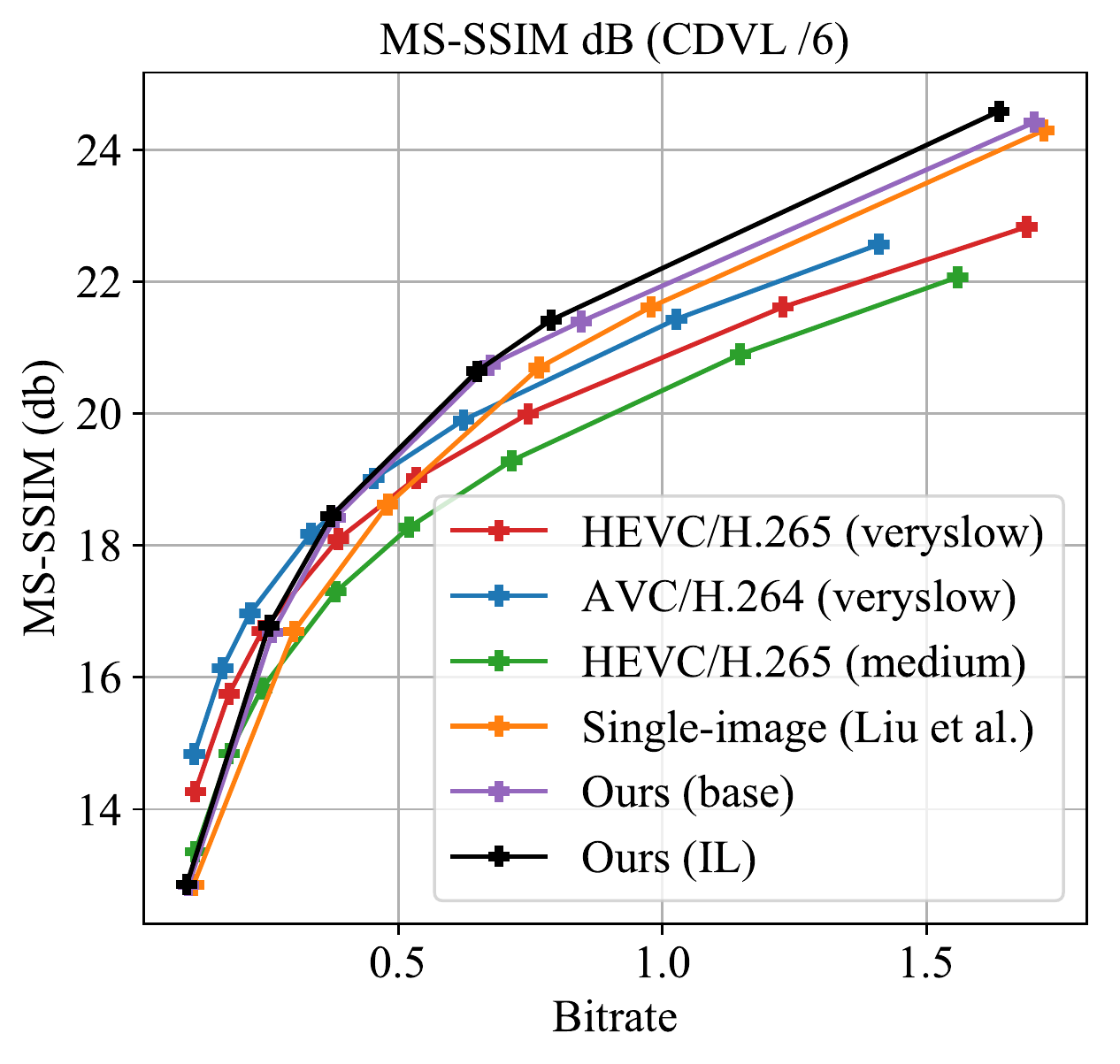} 
	\includegraphics[height=0.23\linewidth]{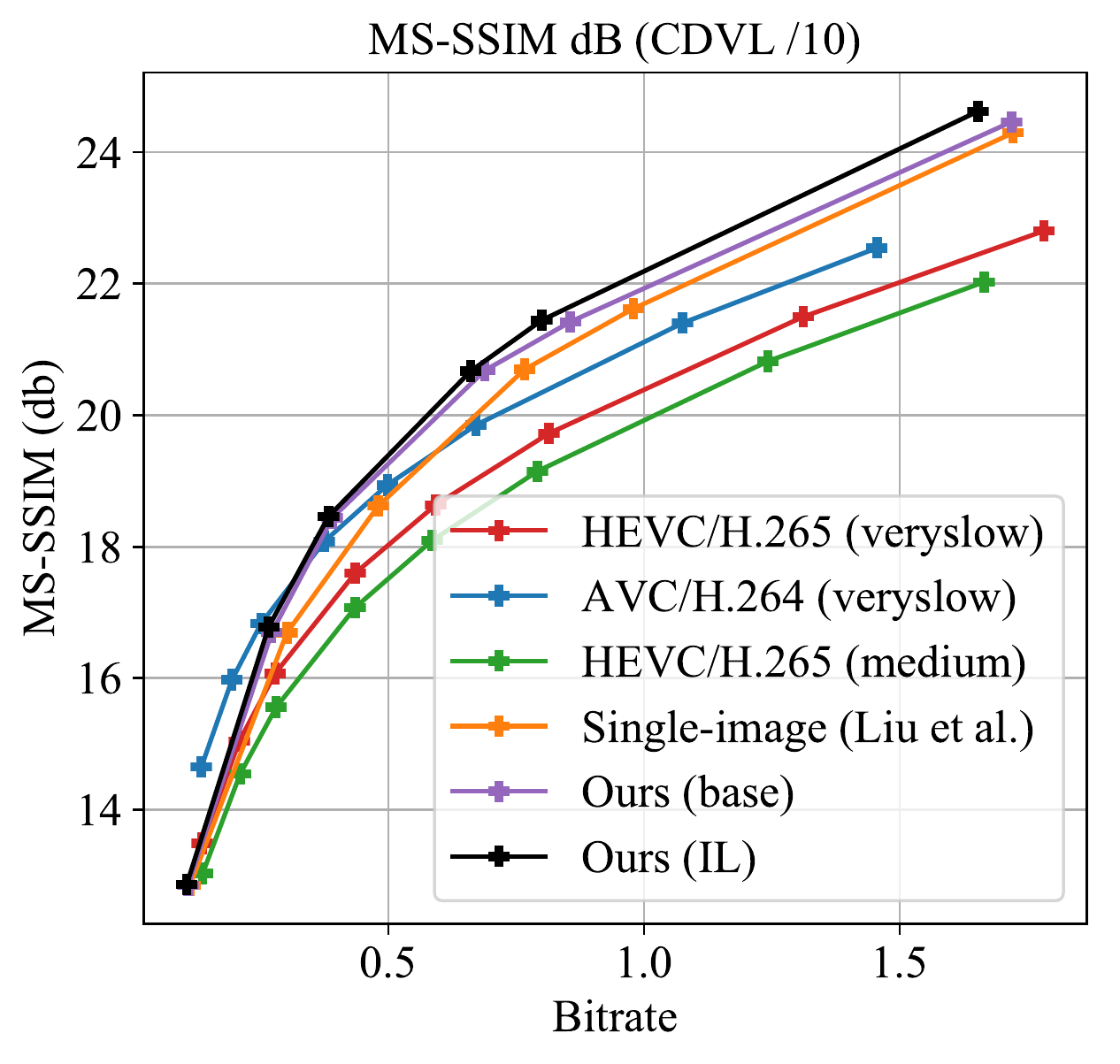} 
	\vspace{-1mm}
	\caption{Plot of our conditional entropy + internal learning adaptations against various baselines for UVG and CDVL. A separate graph is shown for each framerate.
}
	\label{fig:other_fr_results_figure2}
	\vspace{-2mm}
\end{figure*}

\subsection{Varying Framerates on UVG and CDVL}

We can additionally control the framerate by dropping frames for CDVL and UVG. We follow a scheme of keeping 1 out of every $n$ frames, denoted as $/n$. We analyze UVG and CDVL video in 1/3, 1/6, and 1/10 settings. Since all UVG videos are 120 Hz, the corresponding framerates are 40 Hz, 20 Hz, 12 Hz. 

The effects of our conditional entropy model and internal learning, evaluated at different framerates, are shown in separate graphs, in Fig.~\ref{fig:other_fr_results_figure2}. The conditional entropy model is competitive with H.265 at the original framerate for UVG, and outperforms video codecs at lower framerates. In fact, we found that \textit{single-image compression} matches H.265 \textit{veryslow} on lower framerates! We find a similar effect on CDVL at lower framerates as well, where both single-image compression and our approach far outperform H.265 at lower framerates. 


Our base conditional entropy model generally demonstrates a 20\%-50\% reduction of bitrate compared to the single-image model. The effect of internal learning on each frame code provides an additional 10-20\% reduction in bitrate, demonstrating that internal learning of the latent codes during the inference stage provides additional gains.

\subsection{Qualitative Results}
We showcase qualitative outputs of our model vs H.265 and H.264 \textit{veryslow} in Fig. \ref{fig:qual_eval_all}, demonstrating the power of our model on lower framerate video. On 10Hz NorthAmerica, 12Hz UVG video, and 6Hz CDVL video, our model contains big reductions in bitrate compared to video codecs, while producing results that are more even and with fewer artifacts.

%% file: conclusion.tex
\section{Conclusion}

We propose a novel entropy-focused video compression architecture consisting of a base conditional entropy model as well as an internal learning extension. Rather than explicitly transforming information across frames as in prior work, our model aims to model the correlations between each frame code, as well as perform internal learning of each frame code during inference to better optimize this entropy model. We show that our lightweight, entropy-focused method is competitive with prior work and video codecs as well as being much faster and conceptually easier to understand. With internal learning, our approach outperforms H.265 in numerous video settings, especially at higher bitrates and lower framerates. Our adaptations are anchored against single-image compression which is robust against varied framerates, whereas video codecs such as H.265 / H.264 are not. Hence, we demonstrate that such a video compression approach can have wide applicability in a variety of settings.

%% file: supp.tex
\appendix

\input{supp/supp_moredata.tex}

\input{supp/supp_liu.tex}

\input{supp/supp_internal.tex}

\input{supp/supp_codecgop.tex}

\clearpage

%
%

%% file: supp/supp_moredata.tex
\section{Evaluations on Additional Datasets}
We run additional evaluations on two datasets, MCL-JVC \cite{mcl_jvc} and Video Trace Library (VTL) \cite{vtl}. MCL-JVC is a video benchmarking dataset consisting of 1920x1080 video frames ranging from 24-30fps. VTL is a video benchmark dataset consisting of lower resolution video frames (352x288). Due to the wide discrepancy of video lengths in the VTL dataset, we set the maximum video length to 300. We run our standard set of video codec baselines on the datasets (H.265 veryslow, H.265 medium, H.264 veryslow), and set the max Group of Pictures (GoP) size to the length of the video. 

We plot rate-distortion curves on MCL-JVC, and demonstrate that we outperform video codecs while remaining competitive with the state-of-the-art prior work of Djelouah et al. \cite{djelouah_interframe}, which utilizes bi-directional interpolation to depend on both the past and future. 

We also evaluate our approach on VTL, which yields surprising observations. A cursory look at the rate-distortion plots show that our approach far underperforms those of other video codecs, especially on PSNR. However, a closer analysis of the qualitative results show high-frequency artifacts in the source video that our frame encoder does not capture. We discuss these results and offer an explanation for the quantitative discrepancy of our approach with other codecs below. 

\subsection{Rate-Distortion Curve on MCL-JVC}

\begin{figure}
	\centering
	\includegraphics[width=0.4\linewidth]{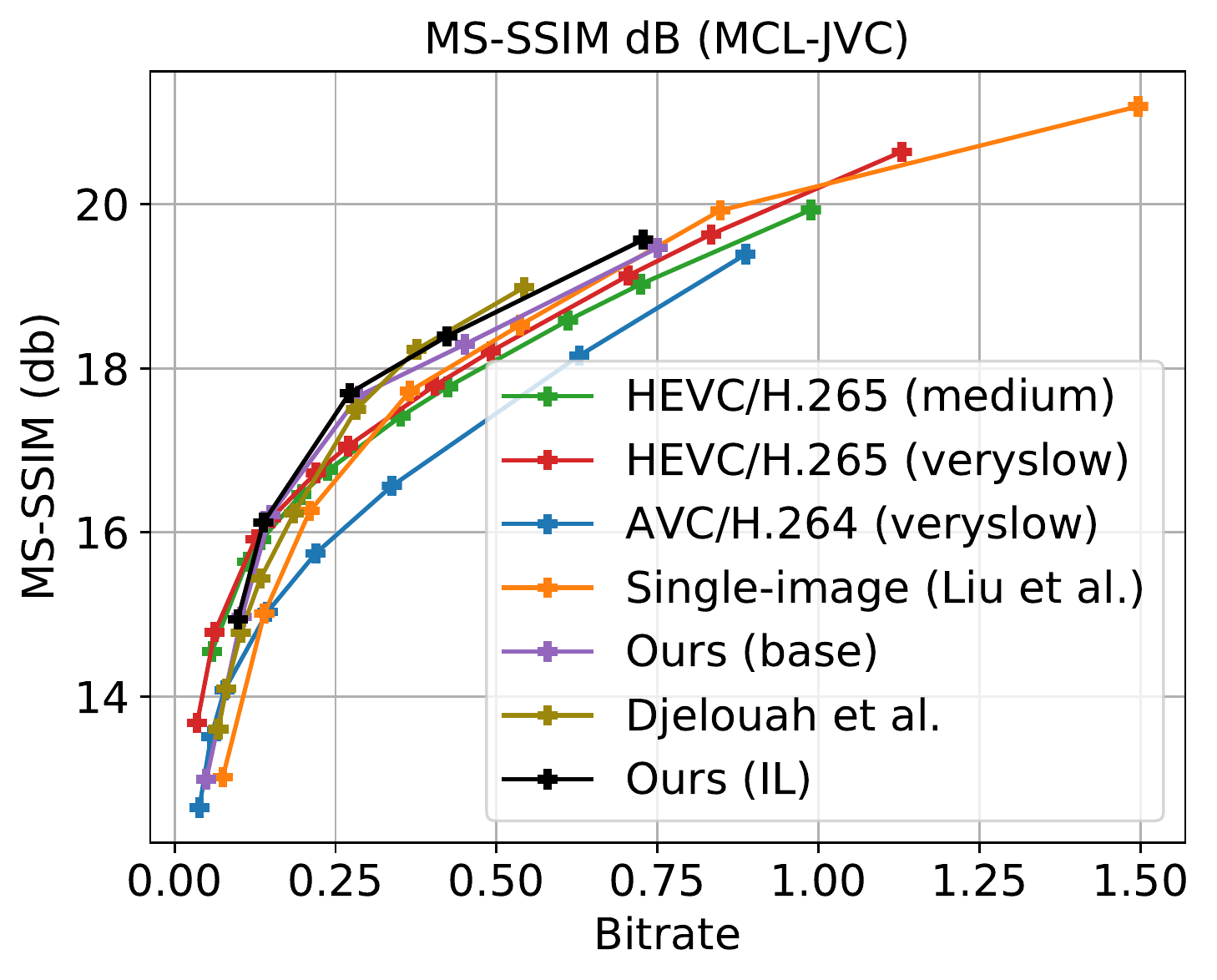}
	\includegraphics[width=0.4\linewidth]{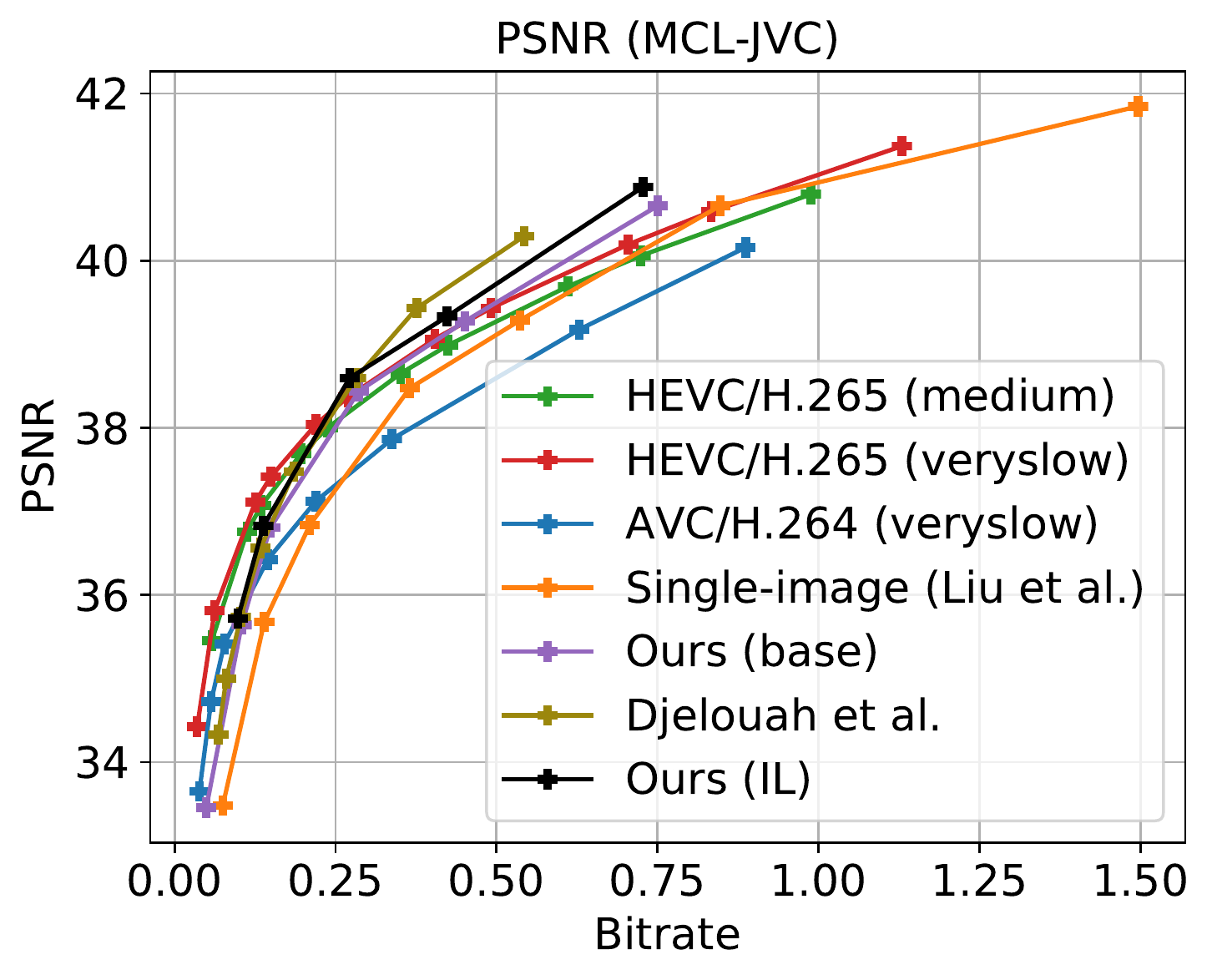}
	\caption{Rate-distortion plot of our approach vs. competing baselines on MCL-JVC.}
	\label{fig:mcl_jvc}
	\vspace{-2mm}
\end{figure}

We plot the rate-distortion curve on MCL-JVC, as shown in Fig. \ref{fig:mcl_jvc}. We observe that our approach outperforms other video codec baselines. Surprisingly, it is also competitive with the state-of-the-art work Djelouah et al. \cite{djelouah_interframe}. This is interesting because Djelouah et al. utilizes a bidirectional model - each intermediate frame depends on a mixture of not only the past, but future frames as well. Meanwhile, in our approach each frame only as a probabilistic dependence on the past frame through the entropy model, as we intend our approach to eventually be applied to an online setting.

\subsection{Analysis of VTL: High-Frequency Information}

\begin{figure}
	\centering
	\includegraphics[width=0.4\linewidth]{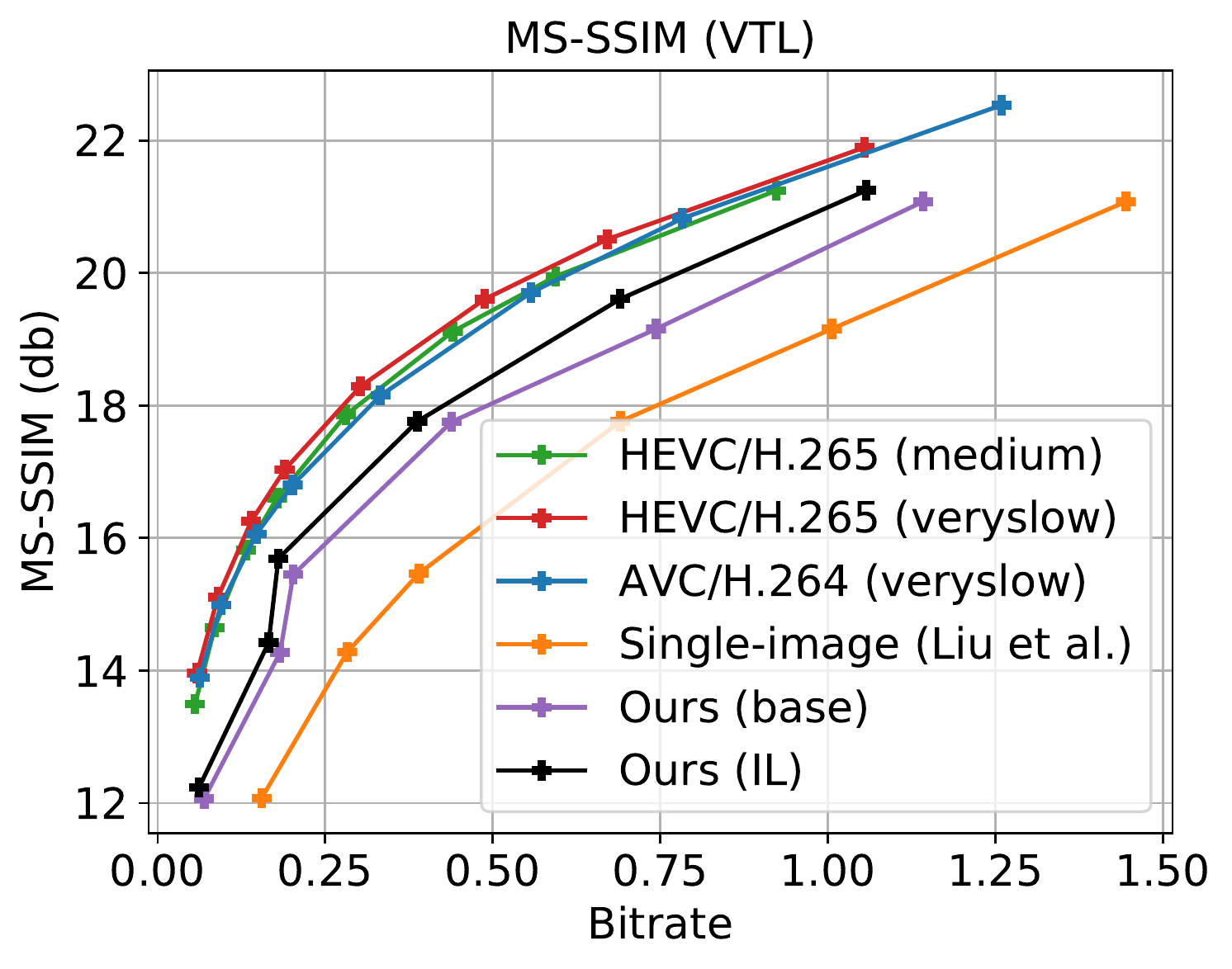}
	\includegraphics[width=0.4\linewidth]{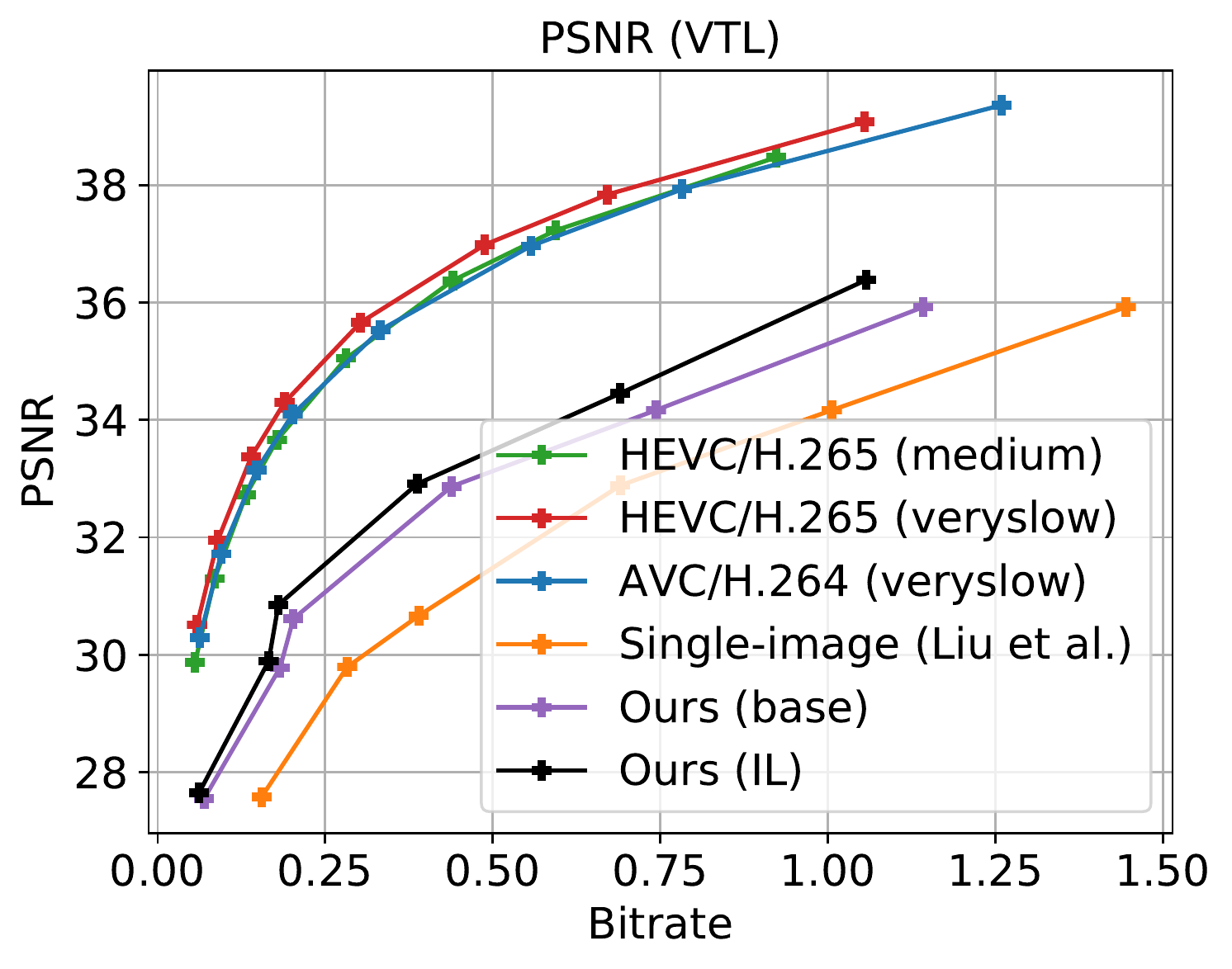}
	\caption{Rate-distortion plot of our approach vs. competing baselines on VTL.}
	\label{fig:vtl}
	\vspace{-2mm}
\end{figure}

\begin{figure}
	\centering
	\includegraphics[width=0.4\linewidth]{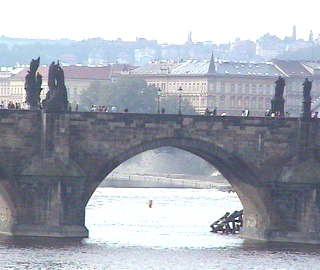}
	\includegraphics[width=0.4\linewidth]{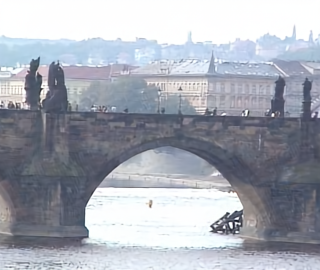} \\
	\includegraphics[width=0.4\linewidth]{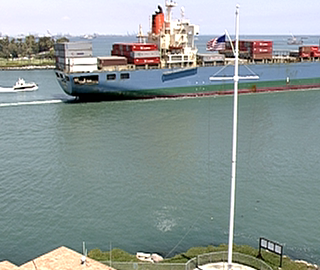}
	\includegraphics[width=0.4\linewidth]{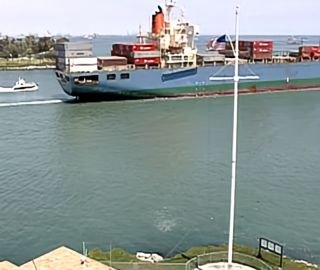}
	\caption{Qualitative comparison of source VTL frames (left) vs. our reconstructions (right)}
	\label{fig:vtl_qual}
	\vspace{-2mm}
\end{figure}

When we initially view the rate-distortion curves in Fig. \ref{fig:vtl}, we see that our approach underperforms video codecs by a large margin, especially on PSNR. In order to understand the quantitative discrepancy, we qualitatively analyze the frames of the source video and reconstructed video, as shown in Fig. \ref{fig:vtl_qual}. We observe that there exist high-frequency information in the source frames, often in the form of artifacts (see the color bands across the bridge), that our frame encoder is not able to capture even at a reasonably high bitrate. 

We offer some hypotheses and discussions of these results. In our approach, each frame is encoded and reconstructed independently with an image encoder / decoder. Our encoder/decoder may contain an inductive bias that might not be able to capture the high-frequency artifacts in the full frame, which explains why they are essentially ``denoised" in the reconstruction. This could be due to the architecture of the encoder/decoder, or due to training the model on a different source dataset (Kinetics).   And because our conditional entropy model only reduces the entropy/bitrate and doesn't improve reconstruction in the same way motion estimation/interpolation does, the performance of our approach is only dependent on the quality of the reconstructions provided by the image encoder/decoder. In the meantime, video codecs utilizing explicit transformations only have to encode full frame information in the I-frames, leaving P-frames and B-frames to only encode the residual high-frequency artifact information. 

There are pros and cons for both approaches. Explicit transformations may reconstruct high-frequency information better in intermediate frames, but can introduce additional artifacts through erroneous/quantized motion estimation and residual coding. Our approach of independent frame encoding guarantees no additional motion artifacts will be introduced, but how well high-frequency information can be encoded depends on the nature of the frame encoder. 


%% file: supp/supp_liu.tex
\section{Additional Architecture Details}

\begin{figure*}
	\centering
	\includegraphics[width=0.9\linewidth]{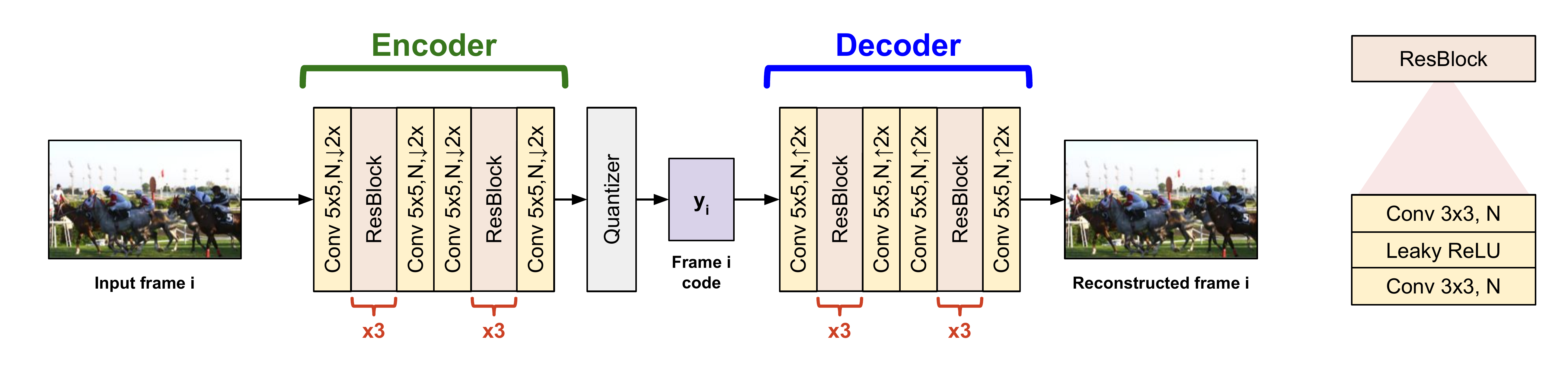}
	\caption{Diagram of our image encoder and decoder model. }
	\label{fig:add_ed_fig}
	\vspace{-2mm}
\end{figure*}

\subsection{Architecture Details of Image Encoder / Decoder}  \label{sec:add_ed_sec}

The architecture of our image encoder and decoder are inspired from that of \cite{liu_nonlocalcomp}. A diagram is shown in Fig. \ref{fig:add_ed_fig}.  The encoder consists of $5 \times 5$ downsampling convolutional layers as well as residual blocks in between. Each residual block consists of a simple sequence of $3 \times 3$ conv, Leaky ReLU, $3 \times 3$ conv layer. The decoder consists of a similar architecture, except each downsampling conv in the encoder is now replaced with a $5 \times 5$ upsampling transposed convolution in the decoder. Let the number of channels in each conv layer be denoted as $N$ (this includes the number of channels of the quantized code). For the lower bitrates, we set $N=192$, and for the higher bitrates, we set $N \in [250,350]$. Since our conditional entropy model imposes its own information bottleneck depending on the $\lambda$ we set in the rate-distortion loss function as well as the target bitrate $R_a$ we want to enforce (see Section 3.3), we intentionally set the number of channels to be higher than necessary so that the channel dimensions themselves do not unnecessarily constrain information. We have not attempted to tune the number of channels for speed performance, though that would certainly provide further gains in speed. 

As mentioned in Section 3.1 in the main paper, we note that we removed all non-local layers as they imposed a large bottleneck of speed and memory usage. 

\subsection{Additional Architecture Details of Hyperprior Encoder / Decoder} 
The architecture of the hyperprior encoder and decoder is listed in Fig. 4 in the main paper. Each residual block referred to in that figure is the same as the residual block defined above in Section \ref{sec:add_ed_sec} and Fig. \ref{fig:add_ed_fig} here in supplementary material. 

In the hyperprior decoder, all feature maps at the spatial resolution of the main image code $\jmb{y}_i$ or lower have $N$ channels, with $N$ being the same as the one defined above. All feature maps at a higher spatial resolution (the ones interspersed with the upsampling/downsampling IGDN/GDN layers) have $M$ channels, with the exception of the highest resolution channel (which has 5 channels). $M$ ranges from $80$ at lower bitrates to $192$ at higher bitrates. Each convolution layer in the hyperprior decoder originally has a kernel size of $5 \times 5$, though we replace the layer with two $3 \times 3$ conv layers for speed gains.

%% file: supp/supp_internal.tex
\begin{figure}
	\centering
	\includegraphics[width=0.4\linewidth]{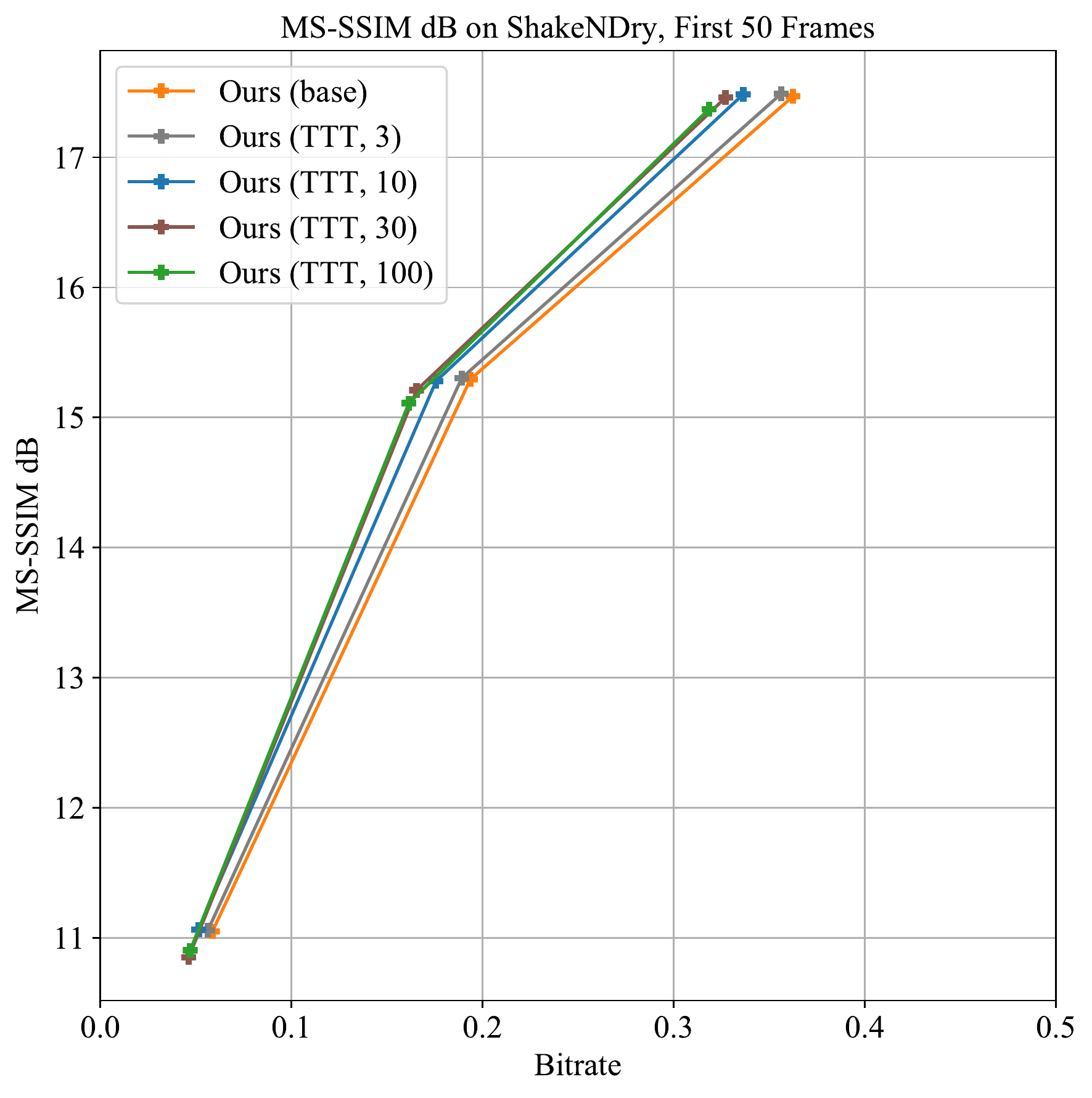}
	\includegraphics[width=0.4\linewidth]{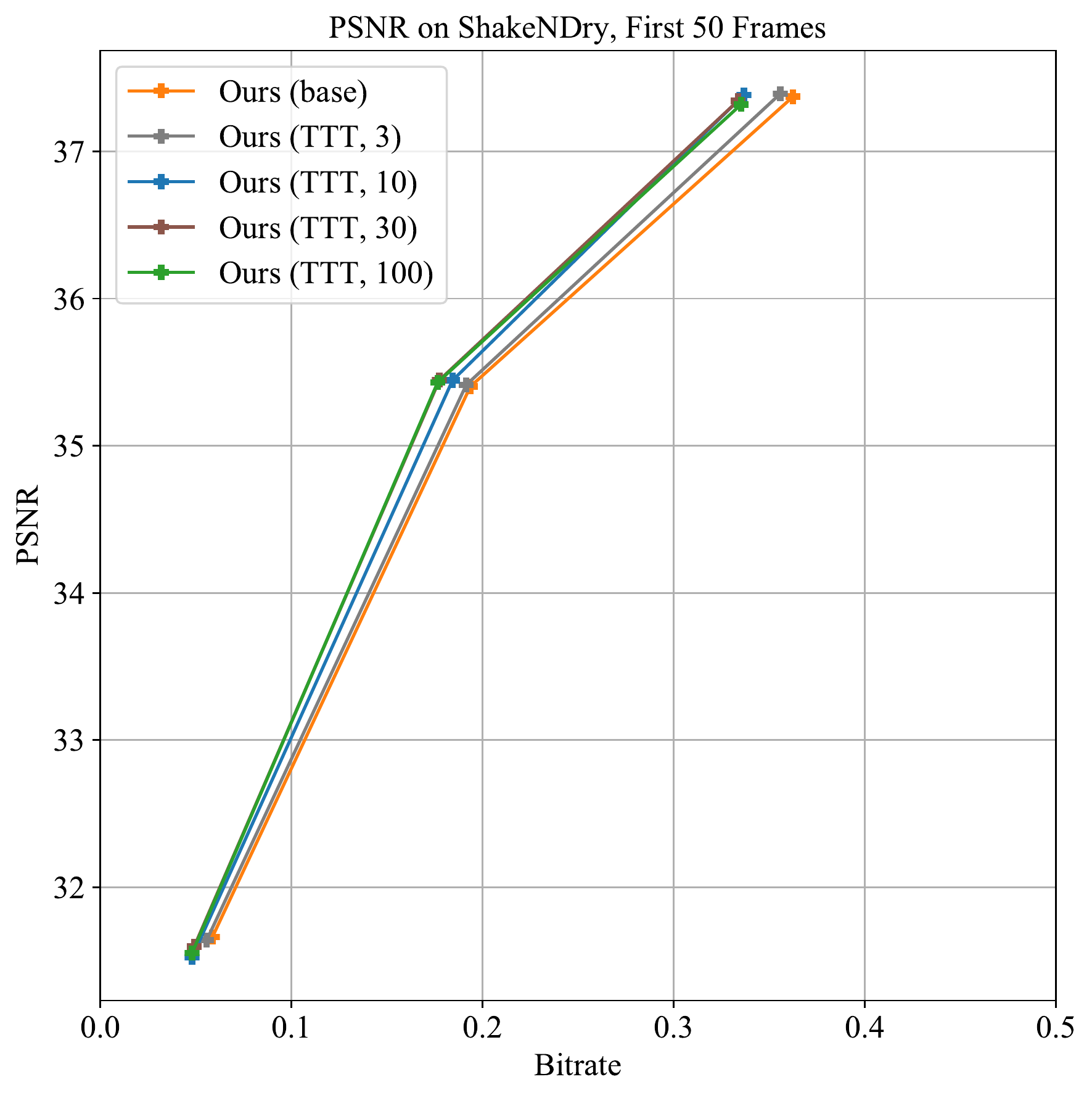}
	\caption{MS-SSIM and PSNR rate-distortion plot on the first 50 frames of the UVG ShakeNDry video sequence of our approach as we vary the number of internal learning steps.}
	\label{fig:internal_learning_steps}
	\vspace{-2mm}
\end{figure}

\section{Internal Learning - Effect of Num. Steps on the Rate-distortion Curve}

We additionally analyze how much the performance of our models improve as we increase the number of gradient steps used in internal learning of our frame latent codes $\jmb{y}_i$ and $\jmb{z}_i$ for each frame $i$. To do this, we evaluate on the first 50 frames of the \verb|ShakeNDry| UVG  video sequence. 

Results are shown in Fig.~\ref{fig:internal_learning_steps} for both MS-SSIM and PSNR, where we plot the rate-distortion curves of our base model as well as with internal learning of 1, 10, 30, 100 gradient steps. Specifically, we use stochastic gradient descent with Nesterov momentum of 0.9 - we decrease the learning rate at higher gradient steps to reduce instability. We can see that in general, increasing the number of gradient steps reduces bitrate and distortion, though performance appears to saturate after 30 gradient steps. In the main paper, we use a fixed number of 10 steps for every test video frame. 

%% file: supp/supp_codecgop.tex
\section{Effect of GoP Size on Codec Performance}
\begin{figure}
	\centering
	\includegraphics[width=0.7\linewidth]{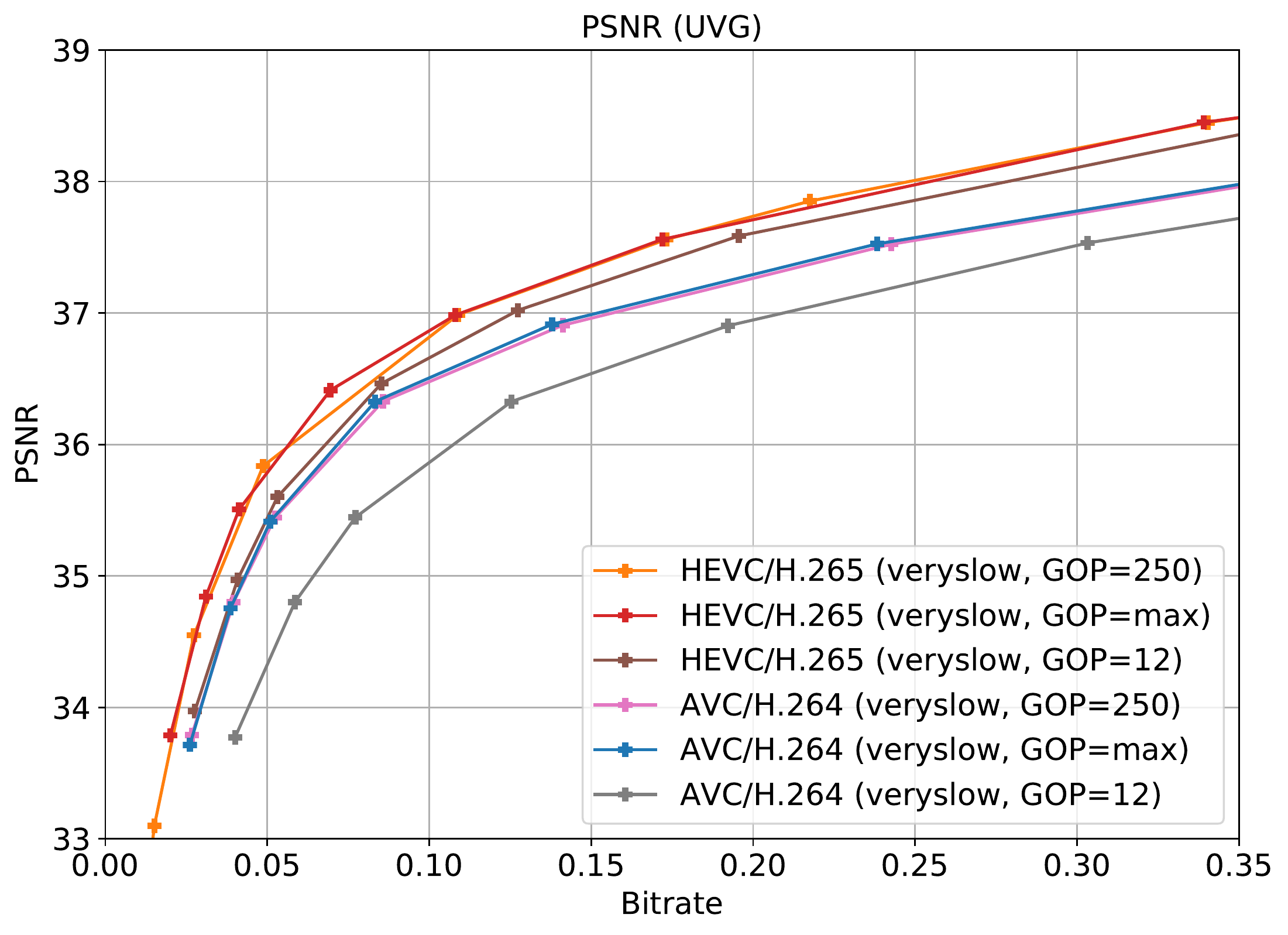}
	\caption{Analysis of GoP size on codec performance in UVG.}
	\label{fig:uvg_gop}
	\vspace{-2mm}
\end{figure}
We benchmark H.265 and H.264 on different Group of Pictures (GoP) settings to analyze how GoP affects codec performance. We adjust the maximum and minimum GoP size by tuning \verb|keyint|/\verb|min-keyint| and \verb|keyint|/\verb|keyint_min| in libx265 and libx264 respectively, in \verb|ffmpeg|. We test using the default \verb|ffmpeg| settings: $(max=250, min=25)$, as well as $(max=12, min=1)$ and $(max=\text{length of video}, min=1)$. The results over UVG video are plotted in Fig. \ref{fig:uvg_gop}. We see that there is a marginal gap between a smaller GoP size of 12 and the default settings; the gap between the default settings and the maximum GoP size is thin.